\documentclass[twocolumn, twocolappendix]{aastex631}

\usepackage{units}
\usepackage{breqn}
\usepackage{bm}
\usepackage{multirow,changepage,mathtools}
\usepackage{xcolor}
\usepackage{longtable}
\newcommand{\Ac}[1]{$\prescript{#1}{89}{\text{Ac}}$}
\newcommand{\Al}[1]{$\prescript{#1}{13}{\text{Al}}$}
\newcommand{\Am}[1]{$\prescript{#1}{95}{\text{Am}}$}

\newcommand{\Bi}[1]{$\prescript{#1}{83}{\text{Bi}}$}
\newcommand{\Bk}[1]{$\prescript{#1}{97}{\text{Bk}}$}

\newcommand{\Cm}[1]{$\prescript{#1}{96}{\text{Cm}}$}

\newcommand{\Fe}[1]{$\prescript{#1}{13}{\text{Fe}}$}
\newcommand{\Fr}[1]{$\prescript{#1}{87}{\text{Fr}}$}
\newcommand{\Hf}[1]{$\prescript{#1}{72}{\text{Hf}}$}
\newcommand{\I}[1]{$\prescript{#1}{53}{\text{I}}$}
\newcommand{\Kr}[1]{$\prescript{#1}{36}{\text{Kr}}$}

\newcommand{\Mo}[1]{$\prescript{#1}{42}{\text{Mo}}$}
\newcommand{\Nb}[1]{$\prescript{#1}{41}{\text{Nb}}$}
\newcommand{\Np}[1]{$\prescript{#1}{93}{\text{Np}}$}

\newcommand{\Pa}[1]{$\prescript{#1}{91}{\text{Pa}}$}
\newcommand{\Pb}[1]{$\prescript{#1}{82}{\text{Pb}}$}

\newcommand{\Pu}[1]{$\prescript{#1}{94}{\text{Pu}}$}
\newcommand{\Ra}[1]{$\prescript{#1}{88}{\text{Ra}}$}
\newcommand{\RRe}[1]{$\prescript{#1}{75}{\text{Re}}$}

\newcommand{\Sb}[1]{$\prescript{#1}{51}{\text{Sb}}$}
\newcommand{\Sn}[1]{$\prescript{#1}{50}{\text{Sn}}$}

\newcommand{\Ta}[1]{$\prescript{#1}{73}{\text{Ta}}$}
\newcommand{\Tc}[1]{$\prescript{#1}{43}{\text{Tc}}$}

\newcommand{\Th}[1]{$\prescript{#1}{90}{\text{Th}}$}
\newcommand{\U}[1]{$\prescript{#1}{92}{\text{U}}$}

\newcommand{\Y}[1]{$\prescript{#1}{39}{\text{Y}}$}

\graphicspath{{./}{figs}}

\begin{document}

\title{Spatial models of \textit{r}-process remnants and their $\gamma$-ray detectability}

\author[0000-0002-9930-3591]{Benjamin Amend}
\affiliation{Center for Nonlinear Studies, Los Alamos National Laboratory, Los Alamos, NM 87545 USA}
\affiliation{Department of Physics and Astronomy, Clemson University, Clemson, SC 29634, USA}

\author[0000-0003-2624-0056]{Christopher L. Fryer}
\affiliation{Center for Nonlinear Studies, Los Alamos National Laboratory, Los Alamos, NM 87545 USA}

\author[0000-0002-9950-9688]{Matthew R. Mumpower}
\affiliation{Theoretical Division, Los Alamos National Laboratory, Los Alamos, NM 87545 USA}
\affiliation{Center for Theoretical Astrophysics, Los Alamos National Laboratory, Los Alamos, NM 87545 USA}

\author[0000-0003-4156-5342]{Oleg Korobkin}
\affiliation{Theoretical Division, Los Alamos National Laboratory, Los Alamos, NM 87545 USA}
\affiliation{Center for Theoretical Astrophysics, Los Alamos National Laboratory, Los Alamos, NM 87545 USA}

\begin{abstract}

We investigate the detectability of gamma-ray emission from long-lived radioactive isotopes in r-process–enriched remnants, focusing on how assumptions about their spatial distribution introduce uncertainty into detection prospects. Using a suite of physically motivated models for the Galactic distribution of kilonova and supernova remnants, we simulate synthetic remnant populations and compute their time-evolving gamma-ray spectra. We then compare these flux predictions to the sensitivity limits of next-generation instruments such as COSI and HEX-P. We find that even under optimistic assumptions, detection probabilities with COSI are extremely low ($\ll 1\%$), and that marginal improvements are only possible with instruments like HEX-P if prior localization is available. The choice of spatial distribution model can lead to more than an order-of-magnitude variation in expected line fluxes at low instrument sensitivities, underscoring the role of spatial modeling as a dominant source of uncertainty. Nevertheless, instrumental capability remains the fundamental bottleneck, and a hybrid mission combining COSI-like sky coverage with HEX-P–level line sensitivity would be required to make detection more probable than not.

\end{abstract}

\keywords{Neutron stars (1108) --- Explosive nucleosynthesis (503) --- R-process (1324) --- Gamma-ray sources (633) --- Gamma-ray lines (631)}

\section{Introduction} \label{sec:intro}

Approximately half of all chemical elements heavier than iron are produced via the rapid neutron capture process, or \textit{r}-process \citep{1957RvMP...29..547B, 1957PASP...69..201C, 1965ApJS...11..121S}. Proposed astrophysical sites for the \textit{r}-process broadly fall into two categories: massive stellar collapses—possibly magnetorotationally or fallback-driven \citep{2019Natur.569..241S, 2015ApJ...810..109N, 2012ApJ...750L..22W, 2021Natur.595..223Y, 1994ApJ...433..229W, 2001ApJ...554..578W,2006ApJ...646L.131F}—and compact binary mergers (CBMs) involving at least one neutron star \citep{1989Natur.340..126E, 1999ApJ...525L.121F, 1974ApJ...192L.145L, 2012MNRAS.426.1940K, 2015MNRAS.448..541J, 2013ApJ...773...78B, 2014ApJ...789L..39W, 1999A&A...341..499R, 2013PhRvD..87b4001H, 2011ApJ...738L..32G, 2008ApJ...679L.117S}. Other exotic sources have also been proposed (see \citet{2021RvMP...93a5002C} for a review). The relative contributions of these sources remain uncertain, and galactic chemical evolution models continue to investigate their roles (e.g. \citet{2004A&A...416..997A, 2014MNRAS.438.2177M, 2019ApJ...875..106C}). These models depend critically on accurate \textit{r}-process yields. Due to limited experimental access to many neutron-rich isotopes, yield predictions often rely on theoretical nuclear models~\citep[e.g,][]{2012MNRAS.426.1940K}. Since 2017, the neutron star-neutron star merger GW170817 
has provided new astrophysical constraints \citep{2017Natur.551...67P, 2017ApJ...848L..27T}, but uncertainties in modeling the associated kilonova light curve (AT2017 gfo) persist \citep{2021ApJ...906...94Z, 2021ApJ...918...44B}, and much of the relevant nuclear physics remains poorly known \citep{2022JPhG...49k0502S}. Additional observations of \textit{r}-process events are therefore essential for constraining both astrophysical and nuclear inputs.

Detecting \textit{r}-process signatures in astrophysical events such as kilonovae and supernovae has proven difficult. Kilonovae are exceedingly rare, and while several candidates have been identified \citep{2022Natur.612..223R, 2013Natur.500..547T, 2013ApJ...774L..23B, 2015ApJ...811L..22J,2017ApJ...843L..34K}, confidently associating them with neutron star mergers and further extracting their r-process content remains challenging. The gravitational wave detection of GW170817 conclusively identified AT2017 gfo as a neutron star merger, but early modeling yielded a wide range of predicted r-process masses \citep{2019ApJ...875..106C}. Continued efforts have shown that light curve modeling suffers from degeneracies involving ejecta geometry, velocity structure, and composition \citep{2021ApJ...910..116K,2023ApJ...958..121T,2024ApJ...961....9F}.  Spectroscopic data have proven similarly inconclusive due to the difficulty in unambiguously identifying atomic absorption features in complex, neutron-rich plasmas \citep{2019Natur.574..497W,2023arXiv230213061T}.

Although the prevailing view is that neutron star mergers dominate r-process production, substantial uncertainties remain in the amount of \textit{r}-process material they produce and their galactic event rate \citep{2018ApJ...855...99C, 2023PhRvX..13a1048A, 2024AnP...53600306R}. As a result, some studies argue that additional sources are required \citep{2019ApJ...875..106C}. Recent simulations suggest that core-collapse supernovae (particularly under conditions of strong neutrino oscillations) may also produce heavy \textit{r}-process nuclei, including material near the second and third peaks \citep{2018ApJ...864..171M,2024RvMP...96b5004V}. For completeness, we include a scenario in which core-collapse supernovae are the dominant r-process source. However, any \textit{r}-process material produced in such events is likely buried deep within the ejecta, making direct observational signatures especially difficult to extract.

One promising approach to overcome these challenges is to shift observational focus from optical/infrared wavelengths to hard X-rays and gamma rays. At these energies, emission is driven directly by radioactive decay 
rather than thermal reprocessing or atomic transitions \citep{2016MNRAS.459...35H, 2019ApJ...872...19L, 2020ApJ...903L...3W, 2021ApJ...919...59C, 2022ApJ...932L...7C, 2024PhRvL.132e2701V}. Many of the decay products of r-process nucleosynthesis are long-lived ($\gtrsim\unit[100]{kyr}$), allowing for potential detection of old remnants within the Milky Way. After the first year, such emission becomes largely independent of ejecta geometry or velocity, and the resulting spectral lines are less easily confused with emission from lighter elements, providing a more direct diagnostic of nucleosynthetic yields.

A few recent studies have begun to explore the detectability of long-lived r-process gamma ray emission from local remnants. \citet{2019ApJ...880...23W} modeled sky locations of kilonova remnants using observed short gamma ray burst (sGRB) offsets, assigning fluxes and comparing to detection thresholds for future MeV\textbf{-range} $\gamma$-ray instruments. While informative, this approach assumed that sGRB offsets from external galaxies accurately reflect the distribution of remnants in the Milky Way, which is not necessarily the case due to differences in host galaxy morphology and formation history \citep[e.g.]{gaspari2023galactic}. Other studies have improved spectral modeling but held distances fixed (e.g., 3–10 kpc), neglecting the full distribution of possible remnant locations \citep{2020ApJ...889..168K, 2022ApJ...933..111T, 2024ApJ...971..143C}.

In this work, we build upon these prior studies by performing the first detailed modeling of the spatial distribution of both kilonova and supernova remnants in the Milky Way. This spatial focus is motivated by two key considerations. First, no prior study has systematically explored how assumptions about remnant locations affect detectability, and our models allow us to quantify this major source of uncertainty. Second, spatial distributions may differ between source classes: while r-process yields are expected to be compositionally similar across mergers and supernovae, their spatial footprints differ due to merger kicks and delay times. If detections become feasible in the future, these spatial differences could offer a means of distinguishing progenitor types.

This paper is organized as follows. In Sec. \ref{sec:spatial-distribution}, we present models for the distribution of galactic supernova and kilonova remnants over the past $\unit[1]{Myr}$. We then discuss the properties of both types of remnants in Sec. \ref{sec:remnant-properties}, and present synthetic spectra for the radioactive decay associated with long-lived \textit{r}-process isotopes in Sec. \ref{sec:gamma-ray-spectra}. We discuss the detection capabilities of the relevant instruments—specifically the COmpton Spectrometer and Imager (COSI) and the proposed High Energy X-ray Probe (HEX-P) and similar instruments—in Sec. \ref{sec:detectability-prospects}, and present detectability prospects for r-process events for each instrument. Finally, we conclude with a summary of our results and a brief discussion on the implications of our predictions in Sec. \ref{sec:summary}.

\section{Spatial Distribution Models for Remnants}
\label{sec:spatial-distribution}

In this section, we present several models for the spatial distribution of r-process-enriched remnants within the Milky Way. These models aim to capture the diversity of environments and evolutionary histories associated with the two dominant sources of r-process enrichment: core-collapse supernovae (SNe) and compact binary mergers (CBMs). The models fall into five categories:

\begin{enumerate}
\item A spiral-arm model, representing the distribution of core-collapse supernova remnants (SNRs), which are assumed to trace ongoing star formation and remain localized near their birth sites.
\item A galactic disk model for kilonova remnants (KNRs), which assumes that all motion remains confined to the galactic plane.
\item A static orbital model, in which CBM remnants evolve ballistically in a fixed galactic potential after receiving randomly oriented natal kicks drawn from a velocity distribution.
\item A dynamic orbital model, which integrates CBM orbits in a time-evolving galactic potential, capturing secular changes to the Milky Way's gravitational field.
\item An isotropic model, in which remnants are displaced radially from their birth sites in a manner consistent with the observed distribution of short gamma-ray burst (sGRB) offsets.
\end{enumerate}

These models differ in their assumptions about progenitor birth environments, the magnitudes and directions of natal kicks, and the role of galactic dynamics in shaping remnant distributions. SNRs are modeled exclusively with the spiral-arm distribution, as their short delay times and low velocities imply minimal displacement from their birth sites. In contrast, KNRs are expected to have more spatially extended distributions, reflecting the unique properties of their compact binary progenitors such as large natal kicks and long inspiral times. Observational support for this distinction comes from comparisons between the host-galaxy offsets of long and short GRBs, which are associated with supernovae and kilonovae, respectively \citep{2016ApJ...817..144B, 2022ApJ...940...56F}. However, these offsets are derived from diverse galaxy types with varying morphologies and evolutionary histories, and are limited to projected (2D) distances rather than true three-dimensional positions. As such, the inferred spatial distributions of sGRBs may not directly translate to Milky Way KNRs. In this work, the disk and isotropic models are used to bracket the plausible range of the spatial extents of KNR distributions, providing limiting cases against which the more physically motivated orbital integration models can be compared.

\subsection{Supernova Remnants - Spiral Arm Model}
\label{subsec:sn-remnants}

While kilonovae are rare, supernovae occur far more frequently, and there are hundreds of documented SNRs in the Milky Way \citep{2019JApA...40...36G, 2022ApJ...940...63R}. Most known SNRs are identified via their radio emission, but our focus on hard X-ray and gamma-ray signatures from long-lived radioactive isotopes implies that the number of potentially detectable remnants could significantly exceed the currently observed population. In particular, remnants may remain undetected if their emission is weak or if they lie in observationally challenging regions, such as near the galactic center. We therefore construct a smooth model of the spatial distribution of core-collapse SNRs to estimate the global population relevant for X-ray and gamma-ray surveys.

\begin{figure}
    \includegraphics[scale=1]{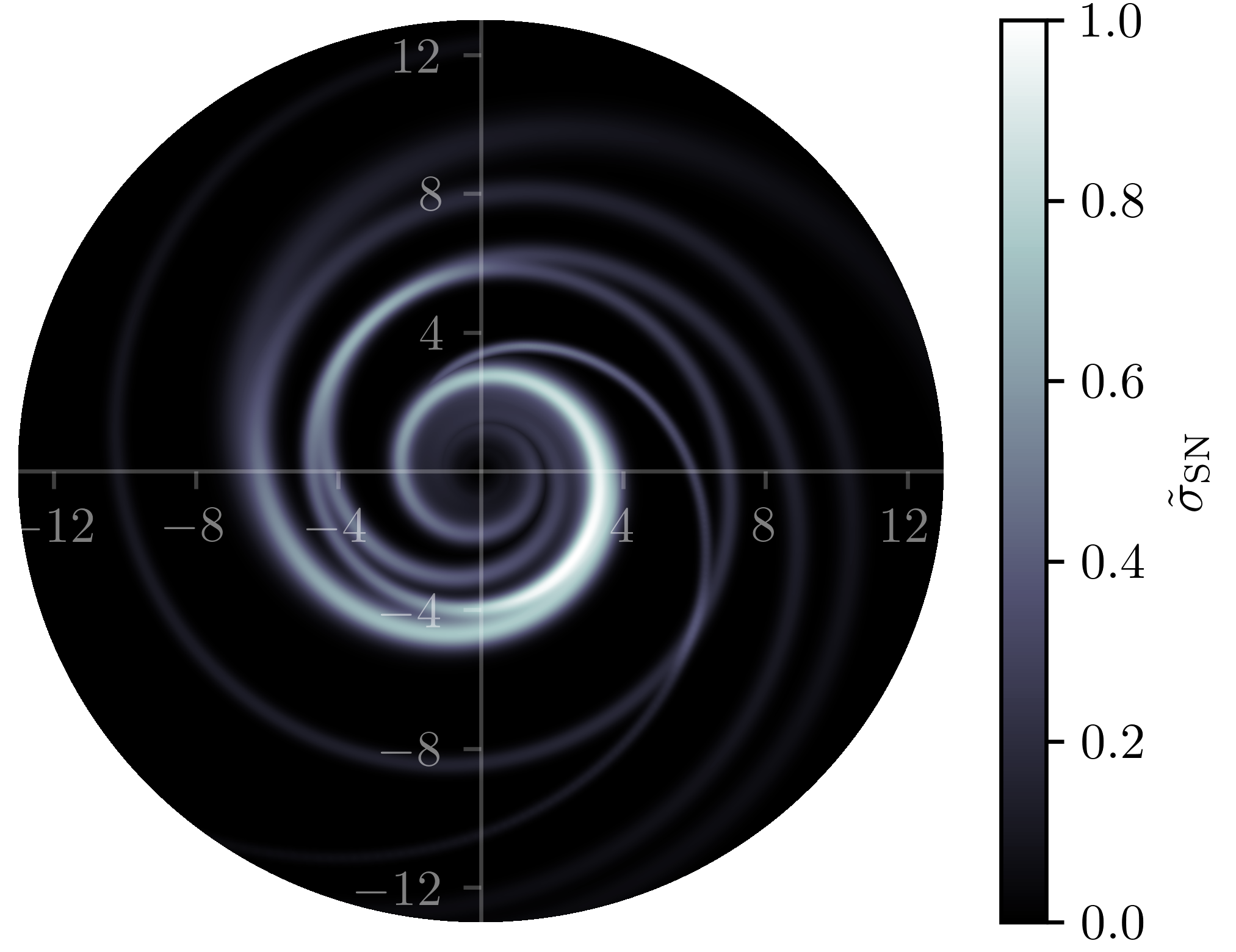}
    \caption{Normalized model surface density of core-collapse supernova locations in the Milky Way (Eq. \ref{eqn:sn-surface-density}). Indicated distances are in units of kpc.
    \label{fig:supernova-distribution}}
\end{figure}

Unlike compact binary mergers, the spatial distributions of SNRs is expected to retain strong correlations with the star-forming regions of the galaxy. In spiral galaxies, core-collapse SNe are concentrated near the spiral arms and follow an overall radial distribution that declines exponentially with galactocentric radius \citep{2005AJ....129.1369P, 1976ApJ...204..519M, 1994PASP..106.1276B, 2009A&A...508.1259H, 1975A&A....44..267B}. We adopt a five-arm logarithmic spiral model based on the fits in \citet{2014ApJ...783..130R}, where the spine of the $i$th spiral arm is described by
\begin{equation}
    R_{\mathrm{spine}}(\beta; R_{\mathrm{ref}}, \beta_{\mathrm{ref}}, \psi) = R_{\mathrm{ref}}e^{-(\beta - \beta_{\mathrm{ref}})\tan{\psi}}.
\end{equation}
Here, $R_{\mathrm{spine}}$ is the galactocentric radius at azimuth $\beta$, $R_{\mathrm{ref}}$ is a reference radius at azimuth $\beta_{\mathrm{ref}}$, and $\psi$ is the pitch angle of the spiral. Motivated by the shapes of the distributions of supernova locations relative to spiral arms in \citet{1976ApJ...204..519M} and \citet{1994PASP..106.1276B}, we model the surface density of the $i$th arm $\sigma_{\mathrm{arm}, i}$ as
\begin{equation}
    \sigma_{\mathrm{arm}, i} \propto \textrm{exp}\lbrace-\left[R-R_{\mathrm{spine}}(\beta; \beta_{\mathrm{ref}, i}, \psi_i)\right]^2 / w_{\mathrm{arm}, i}^2 \rbrace,
\end{equation}
where $w_{\mathrm{arm}}$ is the arm width (a parameter reported in \citet{2014ApJ...783..130R}). For the large-scale radial distribution of supernovae, we adopt an exponential disk profile with scale length $R_s = R_{\mathrm{SN}} / R_{25} \sim 0.29$ \citep{2009A&A...508.1259H}, using $R_{25} \sim \unit[13.4]{kpc}$ for the Milky Way \citep{1998Obs...118..201G}. To account for the observed paucity of supernovae near galactic centers, we introduce a central depletion factor of the form $f(R) = (R/R_s)^2$ for $R < R_s$. The resulting model for the SNR surface density $\sigma_{\mathrm{SN}}$ is then
\begin{equation}
    \sigma_{\mathrm{SN}}(R, \beta) \propto e^{-R / R_s} 
f(R < R_s) \displaystyle\sum_{i=1}^5 \sigma_{\mathrm{arm}, i}.
    \label{eqn:sn-surface-density}
\end{equation}
This distribution is shown in logarithmic color scale in Fig. \ref{fig:supernova-distribution}.

To extend this to a three-dimensional distribution, we introduce a vertical scale height by multiplying the surface density by an exponential factor in $z$. Motivated by the distribution of SNe in the catalog of \citet{2012AdSpR..49.1313F}, we adopt a scale height of $z_s = \unit[100]{pc}$. The full volume density is then
\begin{equation}
    \rho_{\mathrm{SN}}(R, \beta, z) \propto e^{-z / z_s} e^{-R / R_s} f(R < R_s) \displaystyle\sum_{i=1}^5 \sigma_{\mathrm{arm}, i}.
\end{equation}

\begin{figure}
    \includegraphics{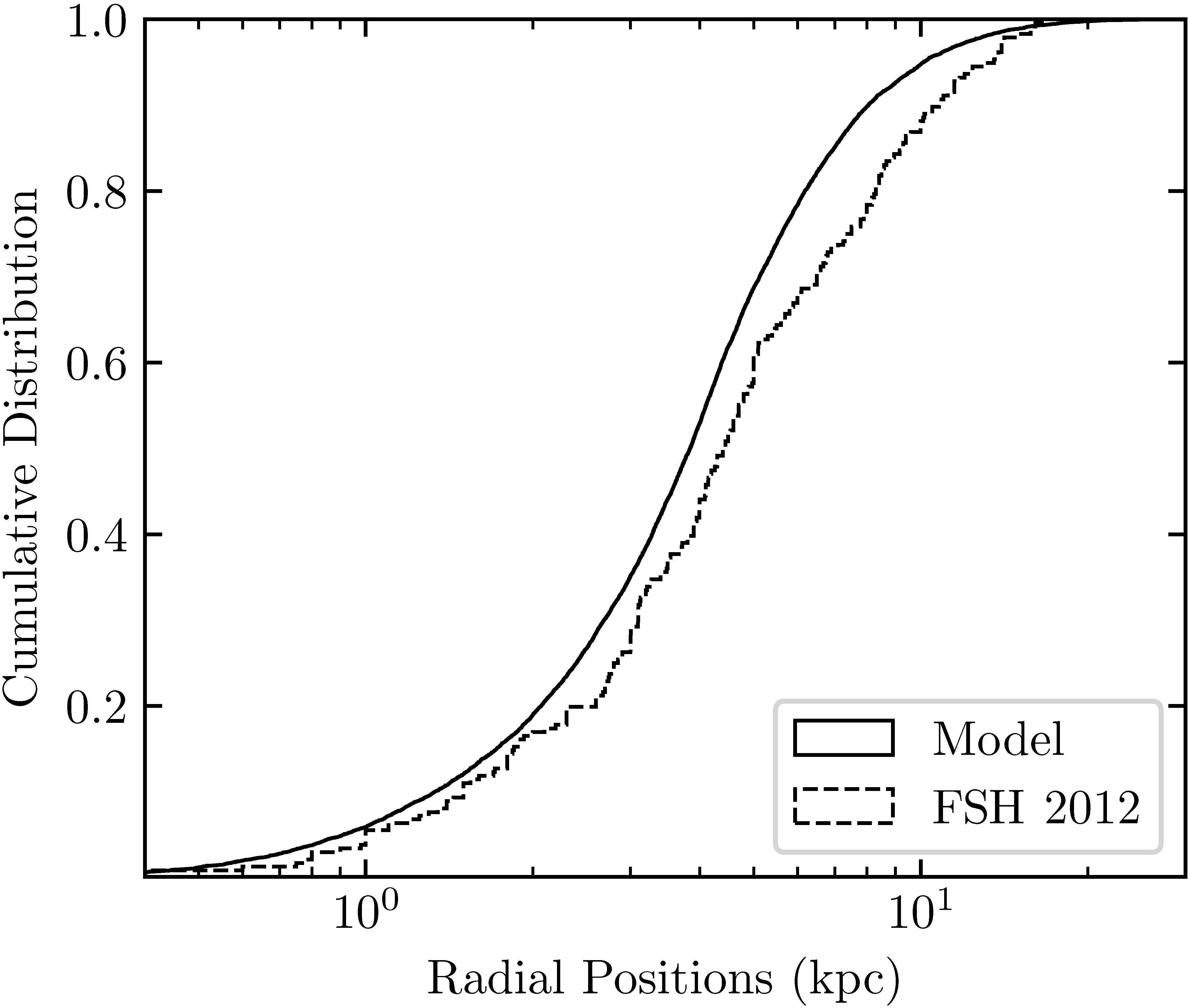}
    \caption{Cumulative distributions of supernova remnant locations relative to the galactic center. The solid curve shows the model prediction, and the dashed curve corresponds to the catalog in \citet{2012AdSpR..49.1313F}.}
    \label{fig:supernova_models}
\end{figure}

In Fig. \ref{fig:supernova_models}, we compare the cumulative radial distribution of SNRs from our model to the observed high-energy SNR catalog compiled in \citet{2012AdSpR..49.1313F}.

This model applies specifically to SNRs, which remain close to their star-forming origins and are not expected to experience significant displacement. In the following subsections, we shift our focus to KNRs, whose compact binary progenitors can travel far from their birth sites before merging, necessitating a broader range of spatial distribution models.

\subsection{Kilonova Remnants - Disk Model}
\label{subsec:disk-model}

If the velocities acquired through supernova kicks are small compared to the galactic escape velocity, or if the binary merger times are short compared to the timescale for significant vertical or radial migration, then most compact binaries will merge within the stellar disk. To represent this regime, we adopt a smooth disk model based on the thickened stellar density profile of \citet{1975PASJ...27..533M}
\begin{equation}
    \rho_*(R, z) = \left( \frac{b_*^2 M_*}{4\pi} \right) \frac{a_* R^2 + \left( 3\sqrt{z^2 + b_*^2} + a_* \right)\left( \sqrt{z^2 + b_*^2} + a \right)^2}{\left[ R^2 + \left( \sqrt{z^2 + b_*^2} + a \right)^2  \right]^{5/2} \left( z^2 + b_*^2 \right)^{3/2}} \, .
    \label{eq:mn_disk}
\end{equation}
The parameters $M_*$, $a_*$, and $b_*$ are chosen to reproduce the Milky Way's present-day stellar mass, local surface density, and local volume density. Specifically, we adopt $M_{*,0} = M_* = \unit[6.08\times 10^{11}]{M_{\odot}}$ from \citet{2015ApJ...806...96L}, and local stellar densities $\Sigma_{*,0} = \unit[33.4]{M_{\odot} pc^{-2}}$ and $\rho_{*,0} = \unit[0.043]{M_{\odot}pc^{-3}}$ from \citet{2015ApJ...814...13M}, which yield scale parameters $a_* = \unit[1.930]{kpc}$ and $b_* = \unit[0.380]{kpc}$. Merger locations are sampled from this density using inverse transform sampling.

\subsection{Kilonova Remnants - Static Orbital Model}
\label{subsec:stat-orb-model}

While disk-confined models represent an important limiting case for KNR spatial distributions, they neglect the wide range of delay times and natal kick velocities expected for compact binary systems. In reality, kilonova progenitors arise from a distribution of birth velocities and merger times, shaped by both stellar evolution and supernova dynamics. As such, a more realistic model should account for this diversity by evolving individual systems throughout the galactic potential. Population synthesis models, combined with orbital integrations, enable us to construct such a physically informed spatial distribution of KNRs.

To this end, we construct a static galactic potential composed of three components: a stellar disk, a gaseous disk, and a dark matter halo. The stellar and gaseous disk potentials adopt the form of the Miyamoto–Nagai profile \citet{1975PASJ...27..533M},
\begin{equation}
    \Phi_{\mathrm{disk}}(R, z) = -\frac{GM}{\sqrt{R^2 + \left( \sqrt{z^2 + b^2} + a \right)^2}} \, ,
\end{equation}
with stellar disk parameters identical to those in Sec. \ref{subsec:disk-model}. For the gas disk, we require a present-day gas mass of $M_{\mathrm{gas},0} = \unit[1.465\times 10^{10}]{M_{\odot}}$, which is somewhat higher than observational estimates (e.g. \citet{2006MNRAS.372.1149F}) but necessary to match local gas properties. Specifically, the parameter choices $a_{\mathrm{gas}} = \unit[5.673]{kpc}$ and $b_{\mathrm{gas}} = \unit[0.167]{kpc}$ reproduce the present-day local gas surface and midplane volume densities, $\Sigma_{\mathrm{gas},0} = \unit[13.7]{M_{\odot} pc^{-2}}$ and $\rho_{\mathrm{gas},0} = \unit[0.041]{M_{\odot} pc^{-3}}$, from \citet{2015ApJ...814...13M}.

We model the dark matter halo with the standard Navarro-Frenk-White (NFW) profile \citep{1996ApJ...462..563N},
\begin{equation}
    \Phi_{\mathrm{DM}}(r) = -\frac{4\pi G\rho_0 r_s^3}{r}\ln{\left( 1 + \frac{r}{r_s} \right)} \, ,
\end{equation}
where $r_s$ is the scale radius and $\rho_0$ is the characteristic density. We calculate $\rho_0$ by requiring that the total halo mass within a cutoff radius of $r_{\mathrm{max}} = \unit[200]{kpc}$ equals the Milky Way's dark matter mass:
\begin{equation}
    \rho_0 = \frac{M_{\mathrm{DM},0}}{4\pi r_s^3}\left[ \ln{\left( \frac{r_s + r_{\mathrm{max}}}{r_s}  \right) - \frac{r_{\mathrm{max}}}{r_s + r_{\mathrm{max}}}} \right]^{-1} \, .
\end{equation}
Subtracting the stellar and gas components from the total virial mass $M_{\mathrm{vir}} = \unit[1.54\times 10^{12}]{M_{\odot}}$ \citep{2019ApJ...873..118W}, we estimate a dark matter mass of $M_{\mathrm{DM},0} = \unit[7.855\times 10^{11}]{M_{\odot}}$, and set $r_s = \unit[7.962]{kpc}$ to match the local dark matter density $\rho_{\mathrm{DM},0} = \unit[0.013]{M_{\odot} pc^{-3}}$ \citep{2015ApJ...814...13M}.

\begin{figure}
    \includegraphics[scale=1.0]{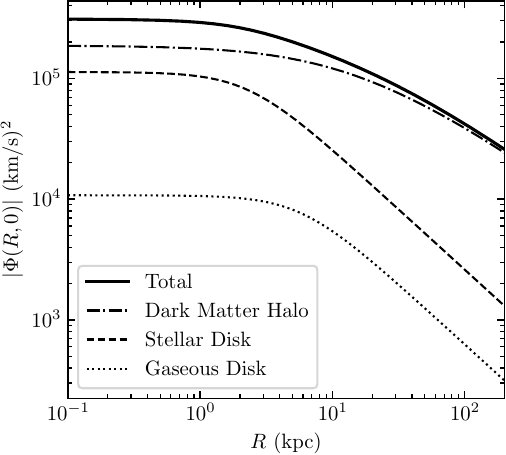}
    \caption{The absolute value of the midplane galactic potential from our static orbital model. The dark matter halo, being the most massive component, largely dominates the potential, with secondary contributions from the stellar and gaseous disks at smaller radii.}
    \label{fig:static_potential}
\end{figure}

The total midplane gravitational potential from all three components is shown in Fig. \ref{fig:static_potential}. The dark matter halo dominates on large scales, while the stellar and gas disks contribute primarily in the inner galaxy, as expected.

To populate this model with merging compact binaries, we first sample birthplaces from the stellar disk density profile (Eq. \ref{eq:mn_disk}), assigning each system a circular orbital velocity given by
\begin{equation}
    v_c(R) = \sqrt{R \frac{d\Phi(R, z=0)}{dR}} \, .
\end{equation}
Each binary is then evolved through the Galactic potential, incorporating both a systemic kick velocity and a delay time before merger.

\begin{figure}
    \includegraphics[scale=1.0]{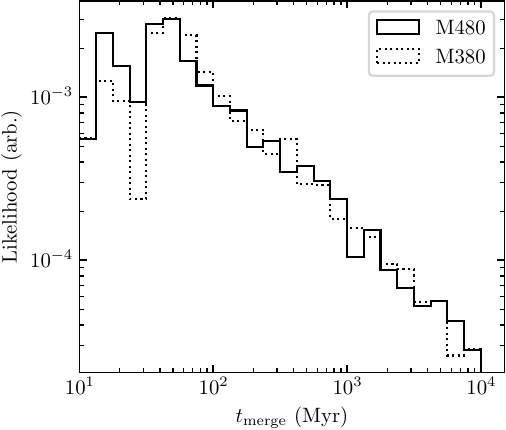}
    \caption{The neutron star-neutron star merger delay time distribution from \texttt{StarTrack} models M380 and M480 for solar metallicity. Both follow an approximate $\sim t_{\mathrm{merge}}^{-1}$ scaling, except for the fastest merging systems.}
    \label{fig:merger_times}
\end{figure}

\begin{figure}
    \includegraphics[scale=1.0]{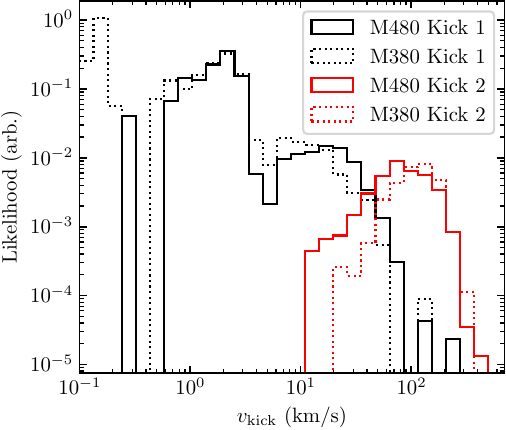}
    \caption{The systemic kick velocity distributions from \texttt{StarTrack} models M380 and M480 for solar metallicity. The first supernova (black) produces a broader range of velocities than the second (red), which tends to favor higher velocities $\sim \unit[100]{km/s}$.}
    \label{fig:kick_velocities}
\end{figure}

\begin{figure*}
    \includegraphics[scale=0.2]{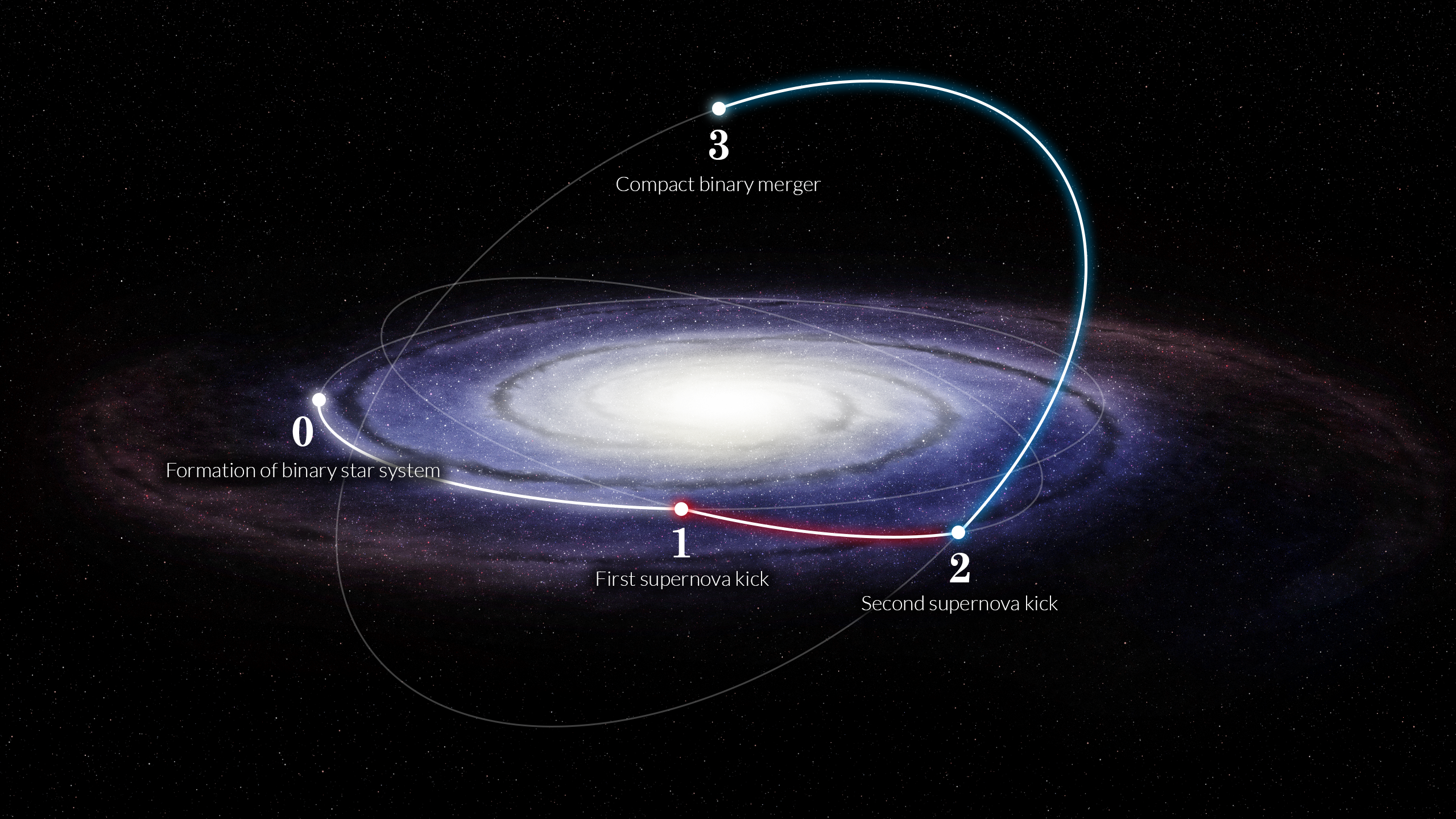}
    \caption{Illustration of a compact binary merger trajectory in the galaxy. The system originates at point \textbf{0} on a circular disk orbit. Supernovae at points \textbf{1} and \textbf{2} impart kicks that incline the orbit and modify the velocity. The system merges at point \textbf{3} after evolving for its sampled delay time.}
    \label{fig:orbit_integrations}
\end{figure*}

We obtain delay times and kick velocities from the models M.380B and M.480B presented in \citet{2021A&A...651A.100O}—produced using the \texttt{StarTrack} population synthesis code \citep{2002ApJ...572..407B, 2008ApJS..174..223B, 2021A&A...651A.100O}. The resulting distributions (Figs. \ref{fig:merger_times} and \ref{fig:kick_velocities}) reflect the broad range of outcomes for compact binary systems. Systems evolve through the galactic potential for the duration of their delay times, during which their orbits are modified by supernova kicks (Fig. \ref{fig:orbit_integrations}). The merger coordinates are recorded at the end of this evolution. Although most systems remain bound to the galaxy, a small fraction ($\sim 1-2\%$) are ejected and merge outside the Milky Way's virial radius.

\subsection{Kilonova Remnants - Dynamic Orbital Model}
\label{subsec:galactic-model}

The most physically motivated model of KNR locations accounts for the time-dependent evolution of both compact binary systems and the Milky Way itself. Unlike the static orbital model, which assumes a fixed galactic potential, this approach recognizes that the merger timescales of neutron star-neutron star and black hole-neutron star binaries can span from tens of Myr to several Gyr \citep{2017ApJ...848L..22B, 2003MNRAS.342.1169V, 2006ApJ...648.1110B, 1999ApJ...526..152F, 1999MNRAS.305..763B}. The subset of systems merging in the recent past (e.g. within the last $\sim\unit[1]{Myr}$) may include a significant number with long delay times. Consequently, their present-day merger locations may be sensitive to the time-varying structure of the galaxy \citep{2018ApJ...865...27W}.

Modeling this evolution is challenging. Galaxy formation and evolution involve a range of physical processes, including gas inflow, star formation, feedback, mergers, and interaction, many of which are difficult to constrain observationally. While some trends can be inferred from present-day properties of the Milky Way itself \citep{1999Natur.402...53H, 2018MNRAS.478..611B, 2018Natur.563...85H, 2019MNRAS.488.1235M}, snapshots of other morphologically similar galaxies at different redshifts (see \citet{2014ARA&A..52..291C} for a detailed review), and simulations ranging in scale from sub-galactic to cosmological \citep{2019ApJ...884...99F}, the specific evolutionary history of the Milky Way remains uncertain. Evidence suggests that the Milky Way has undergone at least six major mergers over its lifetime \citep{2022ApJ...926..107M}, in addition to numerous other dynamical encounters.

Given this complexity, we adopt a simplified time-dependent model that captures the essential growth of the galaxy, modeling time-dependent masses and scale lengths for simple analytic potential profiles. As in the static orbital model, the galactic potential consists of three components: a stellar disk, a gas disk, and a dark matter halo. Here, however, we allow the masses and scale lengths of these components to evolve over time, adopting time-dependent prescriptions from the literature—an approach similar to that used by \citet{2020ApJ...904..190Z}. Full details of this model, including the specific functional forms and various assumptions, are outlined in Appendix \ref{sec:time_dependent_galactic_potential}.

\begin{figure}
    \includegraphics[scale=1.0]{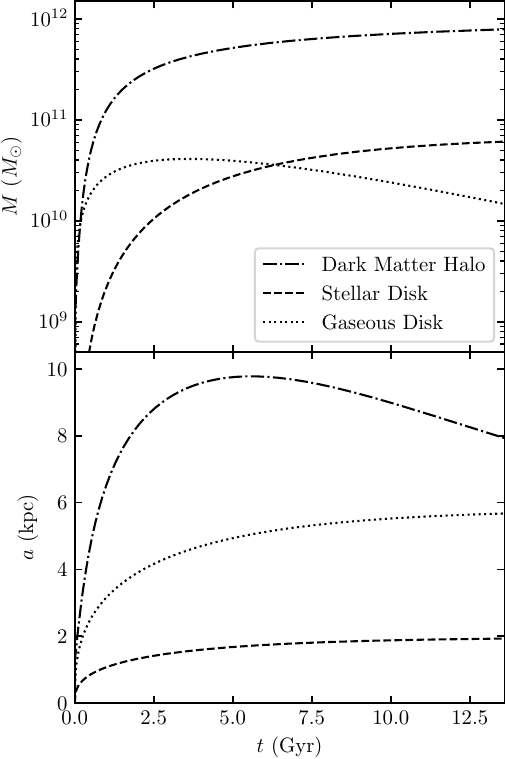}
    \caption{Time evolution of the masses (top) and radial scale lengths (bottom) of the dark matter halo, stellar disk, and gaseous disk. The contraction following the expansion of the dark matter halo is consistent with predictions outlined in \citet{2011ApJ...740..102K}.}
    \label{fig:modelgalaxy}
\end{figure}

The time evolution of each component's mass and radial scale length is shown in Fig. \ref{fig:modelgalaxy}.

Orbits are integrated through this evolving potential using a similar procedure to that in Sec. \ref{subsec:stat-orb-model}.
However, in this dynamic case, the birth time of each system significantly influences its orbital history and final merger location.

To determine these birth times, we convolve a delay-time distribution with a delayed-$\tau$ star formation history. This yields a probability distribution for the formation times of binaries that merge in the present epoch. Additionally, we account for weak metallicity dependence in the population synthesis models by sampling redshift-dependent metallicities following the prescription of \citet{2017ApJ...840...39M}. The full derivation and implementation of these steps are described in Appendix~\ref{sec:binary-population-properties}.

\subsection{Kilonova Remnants - Isotropic Model}
\label{subsec:isotropic-model}

In the regime where compact binary velocities, driven by mass loss and supernova kicks, are large relative to the galactic escape speed, or where merger timescales are long compared to the orbital periods of the binary systems, the spatial distribution of KNRs may become effectively isotropic. In such cases, these remnants no longer trace the disk structure of the galaxy, and their final merger sites can be approximated by a spherically symmetric distribution around the galactic center.

To model this limiting case, we follow the methodology of \citet{2019ApJ...880...23W}, which uses the observed offset distribution of short GRBs from their host galaxies as a proxy for the spatial distribution of compact binary mergers. Here, we adopt the expanded catalog of sGRB host-normalized offsets presented in \citet{2022ApJ...940...56F}. Because the short GRB host galaxies in this catalog span a range of morphologies, including ellipticals, irregulars, and spirals, the resulting offset distribution likely overestimates the typical radial distance for mergers in a Milky-Way-like disk galaxy. However, since this model is intended to serve as a limiting case for extreme spatial dispersion, such overestimation is acceptable and should not detract from its utility.

We draw radial merger distances directly from the observed short GRB offset distribution, using only the host-normalized offsets to reduce dependence on absolute galaxy sizes. These offsets are then rescaled to a fiducial Milky Way radius of $\unit[5.75]{kpc}$ \citep{2024NatAs.tmp..190L}. We sample angular coordinates isotropically from:
\begin{equation}
    \theta = 2\pi u , \hspace{0.25cm} \phi = cos^{-1}{\left( 2v-1 \right)} \hspace{0.5cm} u,v \in [0, 1) \, ,
\end{equation}
where $u$ and $v$ are independent uniform random variables. This yields a fully three-dimensional isotropic distribution of merger sites centered on the Galactic center.

\subsection{Comparison of Kilonova Remnant Model Distributions}
\label{subsec:cbm-distributions}

\begin{figure*}
\begin{center}
    \includegraphics{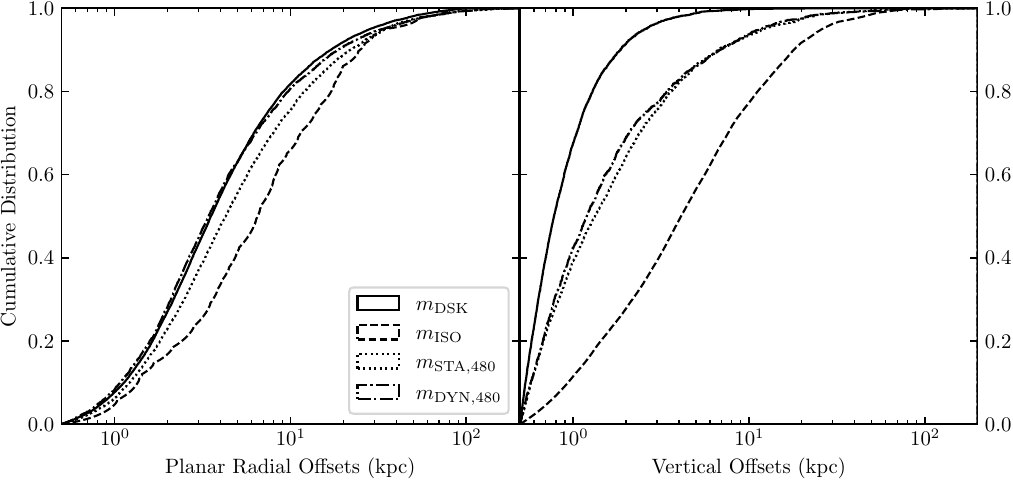}
    \caption{Cumulative distribution functions for planar radial offsets ($R$) and vertical offsets ($z$) for KNRs in the Milky Way for four models: the disk model ($m_{\mathrm{DSK}}$), isotropic model ($m_{\mathrm{ISO}}$), static orbital model ($m_{\mathrm{STA}}$), and dynamic orbital model ($m_{\mathrm{DYN}}$). Results shown are for population synthesis model M.480B; the distributions for M.380B are nearly identical and omitted here for clarity.}
    \label{fig:models}
\end{center}
\end{figure*}

The galactocentric offset distributions for KNRs predicted by each of our four models are shown in Fig. \ref{fig:models}. These cumulative distributions illustrate the range of spatial extents expected for compact binary mergers in the Milky Way, under different assumptions about natal kicks, delay times, and galactic dynamics.

Among these models, the static orbital model produces a mean offset roughly $\sim 25\%$ larger than that of the disk model, consistent with expectations for large spiral galaxies \citep{2006ApJ...648.1110B}. The dynamic orbital model yields slightly smaller offsets, while the isotropic model exhibits significantly broader spatial distributions in both $R$ and $z$.

\section{Remnant Properties}
\label{sec:remnant-properties}

To estimate the detectability of r-process remnants, we must model the physical characteristics of the ejecta from both kilonovae and supernovae. In this section, we describe how we estimate ejecta masses and determine the physical and angular sizes of the resulting remnants.

\subsection{Ejecta Mass}
\label{sec:ejecta-mass}

Neutron star-neutron star (NS-NS) mergers are predicted to eject between $\sim \unit[10^{-4} - 10^{-2}]{M_{\odot}}$ of material, depending on the binary parameters and the neutron star equation of state \citep{2019ARNPS..69...41S, 2013ApJ...773...78B, 2013PhRvD..87b4001H, 2018ApJ...869..130R}. While the \texttt{StarTrack} population synthesis models do not report ejecta masses directly, they do provide the individual neutron star masses for each merging system. We therefore estimate the total ejecta mass using the empirical fitting formula of \citet{2017CQGra..34j5014D}, which expresses the ejecta mass as a function of the component masses and mass ratio.

\begin{figure}
    \includegraphics{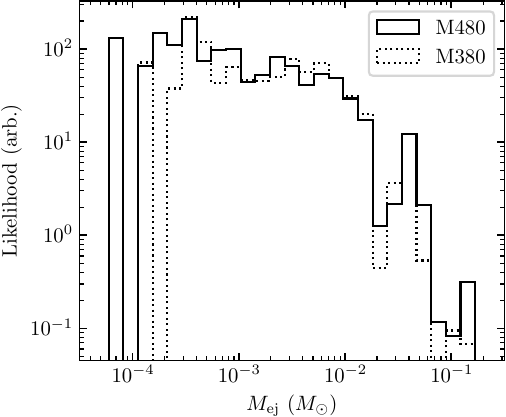}
    \caption{Distribution of neutron star-neutron star merger ejecta masses for progenitors with solar metallicity, computed with the empirical fitting formula of \citet{2017CQGra..34j5014D}, using neutron star masses from \texttt{StarTrack} models M380 and M480.}
    \label{fig:bns_ejecta_mass}
\end{figure}

Fig. \ref{fig:bns_ejecta_mass} shows the resulting ejecta mass distributions for the solar metallicity subset of the M380 and M480 \texttt{StarTrack} models. The typical values cluster around a few times $\unit[10^{-3}]{M_{\odot}}$, but span several orders of magnitude.

In contrast, core-collapse supernovae eject much larger total masses, but only a small fraction of this material is thought to be synthesized via the \textit{r}-process. Estimates for the r-process mass per supernova vary, but are generally in the range $\sim \unit[10^{-6}-10^{-4}]{M_{\odot}}$~\citep{1997ApJ...482..951H,2006ApJ...646L.131F,2007ApJ...659..561M,2018ApJ...864..171M}. For the purposes of this work, we adopt a simple model that uniformly samples r-process masses from this range for each simulated supernova remnant.

\subsection{Remnant Size}
\label{sec:remnant-size}

Both kilonovae and supernovae can be modeled as spherical explosions expanding into a tenuous ambient medium. Initially, the ejecta expand freely at mildly relativistic velocities but eventually decelerate and transition into a Sedov-Taylor-like blast wave phase \citep{1950RSPSA.201..159T}. As the ejecta thrown off from the merger at $v_{\rm ej} \sim 0.1c$ pushes outward into the ISM, a forward shock propagates outward while a reverse shock travels back into the ejecta. These shocks are separated by a contact discontinuity, which traces the outer edge of the ejecta cloud itself.

The extent of the ejecta (not the forward shock) is the physically relevant scale for detecting gamma-ray lines from long-lived r-process isotopes, since the radioactive material is confined to the ejecta. As the ejecta plows through the ISM, it decelerates over the Sedov-Taylor length scale, defined as the point where the swept up ambient mass equals the ejecta mass. This deceleration radius serves as an effective proxy for the final remnant size.

\begin{figure}
    \includegraphics{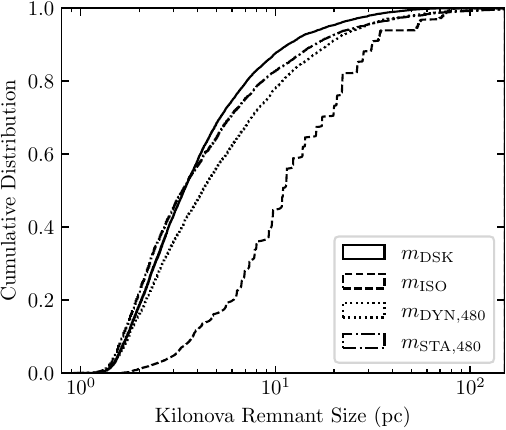}
    \caption{Distribution of KNR sizes for the disk ($m_{\mathrm{DSK}}$), isotropic ($m_{\mathrm{ISO}}$), static orbital ($m_{\mathrm{STA,480}}$), and dynamic orbital ($m_{\mathrm{DYN,480}}$) models. Results are shown only from M.480B since M.380B produces nearly identical distributions. Remnants are largest in the isotropic model due to the prevalence of low-density environments.}
    \label{fig:kn_remnant_size}
\end{figure}

We adopt this deceleration scale to estimate the spatial extent of KNRs. Since it depends only on the ejecta mass and the surrounding medium density~\citep[e.g.][]{2022ApJ...939...59A}, we compute a range of remnant sizes across ambient densities $\sim \unit[10^{-4} - 10^{2}]{cm^{-3}}$, consistent with plausible merger environments identified in \citet{2018ApJ...865...27W}. The resulting kilonova remnant sizes span from a few parsecs in denser regions to tens of parsecs in low-density environments (Fig. \ref{fig:kn_remnant_size}). These correspond to angular sizes of $\lesssim 1^{\circ}$ on the sky, often much smaller, down to arcminutes, at typical distances $\gtrsim \unit[100]{pc}$.

Compared to prior studies (e.g. \citet{2019ApJ...880...23W}), our modeled angular scales are generally smaller, implying that KNRs will often appear as compact or point-like sources. This improves prospects for their identification against the diffuse galactic background.

In contrast to kilonovae, supernovae eject much more material overall but only a small inner fraction is thought to be r-process enriched. The relevant isotopes are likely produced either in the proto-neutron star wind or in fallback accretion onto the neutron star \citep{1997ApJ...482..951H,2006ApJ...646L.131F,2007ApJ...659..561M,2018ApJ...864..171M}, making the \textit{r}-process ejecta more centrally concentrated than, for example, the $^{44}$Ti layer. Observational studies of the spatial distribution of $^{44}$Ti in young supernova remnants \citep[e.g,][]{2014Natur.506..339G} place constraints on the extent of such inner ejecta, which is expected to trace or exceed that of the r-process component. While supernovae have larger ejecta masses and therefore larger overall remnant sizes than kilonovae, they also expand into denser stellar-wind media. In this work, we draw SNR sizes from empirical distributions from galactic catalogs, which indicate angular extents of a few degrees on the sky \citep{2019JApA...40...36G}.

\section{\textit{r}-Process Gamma Ray Spectra}
\label{sec:gamma-ray-spectra}

To estimate the $\gamma$-ray signal coming from the $r$-process, we combine simulations of nucleosynthesis with evaluated nuclear data. 
Prompt nucleosynthesis of the $r$-process as well as the subsequent transmutations of nuclear species on much longer timescales is tracked in the Portable Routines for Integrated nucleoSynthesis Modeling (PRISM) reaction network \citep{Sprouse2021}. 
The theoretical nuclear inputs \citep{Kawano2016, Mumpower2016, Mumpower2020, Mumpower2022} for the network calculation are based on the 2012 version of the Finite-Range Droplet Model \citep{Moller2015, Moller2016}. 
This information is supplemented with measured and evaluated data when applicable \citep{Wang2021, Kondev2021}. 
The trajectory that defines $r$-process conditions is taken from \cite{Rosswog2013} with nuclear self-heating. 
This trajectory produces a main $r$-process and is subjected to robust fission deposition. 
Our main use of these conditions is to produce an abundance pattern that represents a complete $r$-process consisting of the second and third peaks. 
We therefore use these conditions for both kilonovae and supernovae, as the resulting pattern (and subsequent spectra) is largely insensitive to the specific details of the conditions applied given the long observational timescales studied here.
We note that most of the relatively short-lived species undergoing fission do not significantly impact $\gamma$-ray signals on the remnant timescale; hence we do not include the contribution of fission $\gamma$-rays here. 
For a detailed analysis of $\gamma$-rays from fission processes, consult \cite{2020ApJ...903L...3W}.

The radioactive decay spectra associated with long-lived isotopes (relative to the timescale of the $r$ process) are well known and contained in the eighth version of the Evaluated Nuclear Reaction Data Library
(ENDF/B-VIII.0) \citep{Brown2018}. 
Photons emerging from direct transitions in the nucleus and subsequent scattering processes, including X-ray emission from the de-excitation of atomic states and Augur electrons, are provided by this database.
The radioactive decay processes that are considered are $\beta$-decay, $\alpha$-decay and internal transitions between nuclear states. 
The thermal population of isomers is not included in this work. For a comprehensive list of potentially significant gamma-ray emitters on the relevant timescales, see the table in Appendix \ref{sec:gamma-producing-isotopes}.

The abundance information from PRISM is combined with $\gamma$-ray emission data from ENDF to produce time-dependent spectra, 
\begin{equation}
    \label{eqn:spec_fnc}
    S(E,t) = N_A \sum_i \lambda_i Y_i(t) \sum_j I^{\gamma}_j E^\gamma_{j} \delta(E - E^\gamma_{j})
\end{equation}
where $N_A$ is Avogadro’s number, the index $i$ represents the nuclear species with decay rate $\lambda_i$ and abundance, $Y_i$, and the index $j$ represents an associated $\gamma$-ray with intensity $I_j$ and energy $E^\gamma_{j}$, as in \cite{2020ApJ...889..168K}. 

\begin{figure*}
\begin{center}
    \includegraphics[scale=1.0]{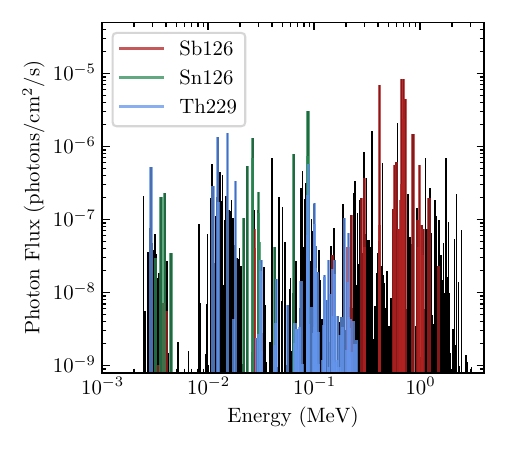}
    \includegraphics[scale=1.0]{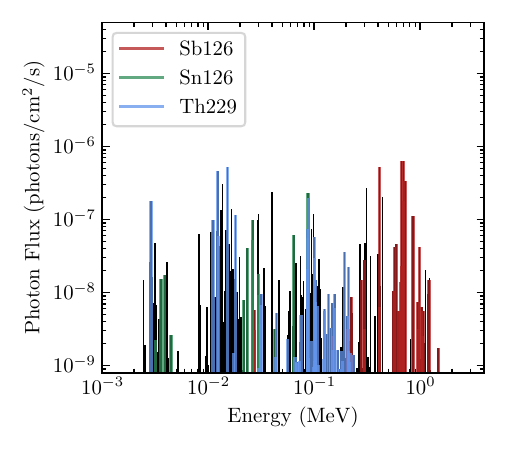}
    \caption{Snapshots of our time-evolving $r$-process spectra at $t=\unit[100]{kyr}$ (left) and $t=\unit[1]{Myr}$ (right) assuming an ejecta mass $M_{\mathrm{ej}}=\unit[0.01]{M_{\odot}}$ and a source distance $D=\unit[1]{kpc}$, with some of the most prominent contributing nuclei indicated.}
    \label{fig:spectra}
\end{center}
\end{figure*}

Equation \ref{eqn:spec_fnc} convolves the population of nuclear species with the decay rate and the intensity of $\gamma$-ray emission. In order to have significant influence on the spectrum, a nuclear species generally needs a balance of these quantities at a given observational time. We plot in Fig. \ref{fig:spectra} an example of our synthetic spectra at $t=\unit[100]{kyr}$ and $t=\unit[1]{Myr}$ to showcase this point. The $\sim\unit[]{MeV}$ region is dominated by lines from $^{126}$Sb \citep{Qian1998}, while the $\sim\unit[]{keV}-\unit[100]{keV}$ region consists of a series of prominent lines from $^{126}$Sn and $^{229}$Th. This apparent contribution from $^{229}$Th at late times arises primarily from the $\alpha$-decay of $^{233}$U, which feeds into the $^{229}$Th decay chain. Although $^{229}$Th itself has a relatively short half-life ($\sim \unit[7.9\times10^{3}]{yr}$), the long half-life of its parent $^{233}$U ($\sim \unit[1.6\times10^5]{yr}$) allows for a sustained presence of $^{229}$Th in the ejecta at timescales of up to $\sim \unit[10^6]{yr}$, contributing to the gamma-ray signal at late times.

There has been some recent interest in the 2.6 MeV gamma-ray line from the decay of $^{208}$Tl–while this line is not included in our synthetic spectra due to its extremely short half-life ($\sim$3 minutes), it is an interesting case to consider. This nucleus resides at the end of several actinide decay chains, such as that of $^{232}$Th, which has a half-life $\sim\unit[]{Gyr}$, meaning it could in principle trace \textit{r}-process nucleosynthesis sites. However, given the observational timescales discussed in this paper, the detectability of this line remains highly uncertain, and is left for future work.

Previous studies have applied varying degrees of Doppler broadening to their spectra based on the assumption that kilonova remnants expand at high speeds for extended periods. However, as we have emphasized, the r-process material of interest remains within the ejecta, confined inside the contact discontinuity, rather than being dispersed throughout the ambient medium swept up by the forward shock. By the time the forward shock enters the self-similar Sedov-Taylor phase, the ejecta has already undergone significant deceleration. On the timescales relevant to our study, the ejecta should be moving at velocities comparable to the nearly static ISM. In this regime, the dominant source of Doppler broadening is thermal motion, which, for standard ISM conditions, leads to a negligible broadening of $\boldsymbol{\ll 1\%}$. This is far below the instrumental broadening already accounted for in sensitivity curves for instruments such as COSI. Consequently, we argue that Doppler broadening in this regime is not as significant as previously suggested in the literature and is effectively negligible for our detectability estimates.

\section{Detectability Prospects}
\label{sec:detectability-prospects}

Detecting gamma-ray emission from long-lived r-process isotopes requires instruments with high spectral resolution, low background, and narrow-line sensitivity in the $\sim\unit[10]{keV}$ to few $\unit[]{MeV}$ energy range. Angular resolution and field of view are also critical: remnants must be sufficiently compact to be distinguished from the diffuse galactic background, and a wide field of view increases the chances of capturing these rare, near-randomly distributed events. Based on these criteria, we focus on two representative instruments: COSI, which offers wide-field survey capability with moderate angular resolution and excellent line sensitivity in the $\sim\unit[100]{keV}$ to $\sim$ a few $\unit[]{MeV}$ regime; and a HEX-P-like hard X-ray instrument, which would offer much higher sensitivity at $\sim$ tens of $\unit[]{keV}$ but a narrower field of view.

The spatial distribution of these remnants determines both the distances and sky locations of potential sources, which in turn affect flux attenuation and whether the source lies in a region favorable for detection. By applying this framework across multiple KNR and SNR distribution models, we directly quantify how assumptions about progenitor kinematics, natal kicks, and galactic evolution influence the likelihood of detecting long-lived r-process emission. This allows us to isolate the role of remnant distribution as a leading source of uncertainty in detection prospects.

To evaluate the observational prospects for detecting long-lived gamma-ray emission from r-process remnants, we simulate synthetic populations of events and compare their predicted photon fluxes to the sensitivity thresholds of representative high-energy observatories. Each simulated remnant is assigned a sky position and distance drawn from one of the spatial distribution models described in Sec. \ref{sec:spatial-distribution}, along with a corresponding age sampled uniformly over the past $\unit[10]{kyr}-\unit[1]{Myr}$. Remnant ejecta masses are assigned according to the progenitor class, as described in Sec. \ref{sec:ejecta-mass}, and the corresponding radioactive decay spectra are computed using the nuclear reaction network and $\gamma$-ray line libraries outlined in Sec. \ref{sec:gamma-ray-spectra}.

Given the ejecta mass and remnant age, we compute the total line-integrated photon flux at Earth,

\begin{equation}
    F_{\gamma}(E,t) = \frac{L_{\gamma}(E,t)}{4\pi D^2}\,,
\end{equation}

where $L_{\gamma}(E,t)$ is the line luminosity at time $t$ after the explosion and $D$ is the remnant's distance from Earth. Because the emission is composed of narrow lines at known energies, we evaluate detectability by comparing this flux to the $3\sigma$ narrow-line sensitivity curves for each instrument under consideration.

We also impose a conservative angular size cut to account for the diminished detectability of extended sources. Specifically, we require that a remnant’s angular diameter on the sky be less than twice the instrument’s full-width at half-maximum (FWHM) angular resolution. This approximates the effective size beyond which spatial confusion with diffuse background emission becomes significant and the source may no longer appear point-like. For kilonova remnants (KNRs), which typically span angular sizes $\lesssim 1^{\circ}$, this criterion is rarely limiting. Supernova remnants (SNRs) are more frequently excluded due to their larger extent on the sky.

By repeating this procedure across $10^6$ realizations for each spatial model, we obtain statistically robust predictions for the number of remnants expected to be detectable by a given instrument. In the following sections, we apply this framework to evaluate detection prospects for COSI and HEX-P-like missions, and explore how detectability improves under hypothetical future instruments with enhanced sensitivity.

\subsection{COSI Results}
\label{subsec:cosi}
The COmpton Spectrometer and Imager (COSI) is a next-generation Compton telescope designed to provide high-resolution imaging and spectroscopy of gamma ray emission in the $\sim \unit[0.2-5]{MeV}$ range using advanced germanium detectors \citep{2019BAAS...51g..98T, 2023arXiv230812362T}. It combines moderate angular resolution (2.1$^{\circ}$-4.5$^{\circ}$ FWHM) with excellent spectral resolution ($\sim 0.8\%-1.1\%$ FWHM) and all-sky survey capability on daily timescales. Its narrow-line sensitivity in the MeV band, estimated at $\sim 3.0\times10^{-6} - 1.2\times10^{-5}\unit{photons \cdot cm^{-2} \cdot s^{-1}}$, is roughly an order of magnitude better than that of INTEGRAL SPI in the same energy range. Together, these attributes make COSI a compelling instrument for detecting $\gamma$-ray lines from the radioactive decay of long-lived $r$-process isotopes in compact remnants. Moreover, its wide field of view is particularly advantageous in this study, as the precise locations of KNRs are not known a priori.

Tab. \ref{tab:cosi-prospects} presents our predictions for the number $N$ of detectable remnants over a 24-month COSI survey, based on synthetic populations drawn from each spatial model. We adopt a kilonova event rate of $\mathcal{R}_{\mathrm{KN}} = \unit[10^{-5}]{yr^{-1}}$, near the lower end of recent LIGO/Virgo estimates for Milky-Way-like galaxies \citep{2023PhRvX..13a1048A}. Even under the most optimistic spatial assumptions, predicted detection probabilities are exceedingly small, at the level of $\boldsymbol{\sim$ $0.01\%}$. The spiral-arm constrained SNR models yield similar results for a supernova rate of $\mathcal{R}_{\mathrm{SN}} = \unit[10^{-2}]{yr^{-1}}$, provided the remnants are compact enough to fall within COSI's angular resolution limits.

\begin{deluxetable}{lcc}
\tablewidth{0pt}
\tablecaption{Number of KNRs and SNRs detectable by COSI over a 24-month survey with $3\sigma$ limits assuming event rates of $\mathcal{R}_{\mathrm{KN}}=\unit[10^{-5}]{yr^{-1}}$ and $\mathcal{R}_{\mathrm{SN}} = \unit[10^{-2}]{yr^{-1}}$.}
\tablehead{
    \colhead{Model} &
    \colhead{$N$}
}
\startdata
\multicolumn{2}{c}{\textbf{Kilonova Remnants}} \\ 
Disk & $6.257 \times 10^{-4}$\\
Static Orbital M380 & $3.705 \times 10^{-4}$\\
Static Orbital M480 & $5.385 \times 10^{-4}$\\
Dynamic Orbital M380 & $3.716 \times 10^{-4}$\\
Dynamic Orbital M480 & $1.797 \times 10^{-4}$\\
Isotropic & $4.947 \times 10^{-5}$\\
\hline
\multicolumn{2}{c}{\textbf{Supernova Remnants}} \\ 
Spiral Arm Distribution & $1.170 \times 10^{-4}$\\
\enddata
\end{deluxetable}
\label{tab:cosi-prospects}

These results indicate that even under the most favorable assumptions about remnant locations, the probability of detecting a single KNR or SNR with COSI is vanishingly small. In this regime, differences between spatial models have little practical impact on detectability, as all plausible distributions yield similarly low event counts. Still, these spatial models remain valuable for identifying where remnants are most likely to be found, and for guiding the development of future observational strategies. Substantial improvements in instrument sensitivity will be necessary before differences between models can translate into observable consequences.

Given these extremely low detection probabilities, it is instructive to explore how detectability improves for a hypothetical future instrument with the same observational strategy and design characteristics as COSI, but with enhanced sensitivity. We define such a COSI-like instrument as having comparable angular resolution, energy bandpass, and sky coverage, but with a sensitivity improved by a multiplicative factor $f$.

We report the fraction of detectable remnants as a function of sensitivity enhancement, evaluated over a $\unit[1]{Myr}$ observation window for each spatial model. Importantly, we do not assume a fixed event rate in this analysis. Instead, detection fractions can be scaled to any assumed event rate $\mathcal{R}$ via
\begin{equation}
    N = f_{\mathrm{det}} \times (\mathcal{R} \times \unit[10^6]{yr})\,,
\end{equation}
where $f_{\mathrm{det}}$ is the fraction of simulated remnants detected at a given sensitivity, and $\mathcal{R}$ is the event rate. This approach avoids overcommitting to a single, uncertain value for $\mathcal{R}$, and makes the results broadly applicable across a range of plausible merger rates and model assumptions.

\begin{figure}
    \includegraphics{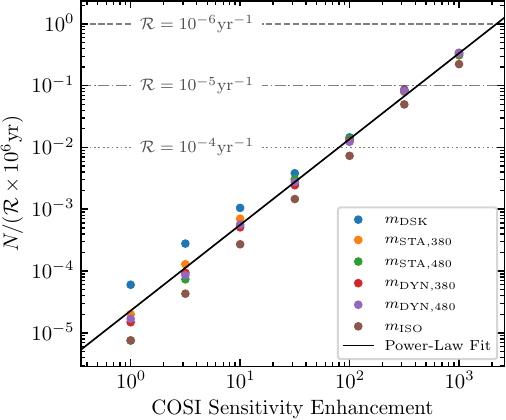}
    \caption{Fraction of detectable kilonova remnants $f_{\mathrm{det}}=N/(\mathcal{R}\times \unit[10^6]{yr})$ vs. sensitivity enhancement factor $f$ for a COSI-like instrument, across different spatial models. Detection fraction scales approximately as $f_{\mathrm{det}} \propto f^{1.388}$. Horizontal dashed lines mark detection fractions of $f_{\mathrm{det}}=10^0, 10^{-1},$ and $10^{-2}$, which yield one expected detection for event rates of $\unit[10^{-6}]{yr^{-1}}$, $\unit[10^{-5}]{yr^{-1}}$, and $\unit[10^{-4}]{yr^{-1}}$, respectively. Differences between spatial models are most pronounced in the low-sensitivity regime, where only the nearest remnants contribute.}
    \label{fig:enhancements}
\end{figure}

We find that this detection fraction scales approximately as a power law with sensitivity enhancement $f$ (see Fig. \ref{fig:enhancements}):
\begin{equation}
    \frac{N}{\mathcal{R}\times\unit[10^6]{yr}} = 2.388\times 10^{-5}f^{1.388}\,,
\end{equation}
where the left-hand side is the detection fraction $f_{\mathrm{det}}$. Fig. \ref{fig:enhancements} shows this scaling for each spatial model, reflecting modest differences between them.

Importantly, spatial model differences are most pronounced in the low-detection regime, where realistic instruments currently operate. At these low sensitivities, only the closest and most favorably positioned remnants could be detectable. As a result, compact spatial distributions (such as the disk or low-kick dynamic orbital models) yield higher detection fractions than more extended or isotropic distributions. E.g. for $f < 100$, there is $\gtrsim$ an order of magnitude spread in the detection fractions between the boundary cases of the disk and isotropic spatial models. Of course, as $f$ increases further and a more significant fraction of remnants in each model becomes detectable, this distinction between spatial models begins to diminish.

\subsection{Targeted Observations in X-Rays}
\label{subsec:hexp}
X-ray detectors are generally far more sensitive than gamma-ray instruments, but they also contend with significantly higher background levels, particularly in crowded regions such as the galactic plane. However, photons from radioactive decay are limited to well-defined line energies. By limiting the search to these line energies, it becomes possible to detect signals well below the broader background continuum. This approach is particularly effective for KNRs, where line emission from long-lived r-process isotopes can persist for $\gtrsim\unit[10^5]{yr}$.

\begin{figure}
    \includegraphics[scale=1.0]{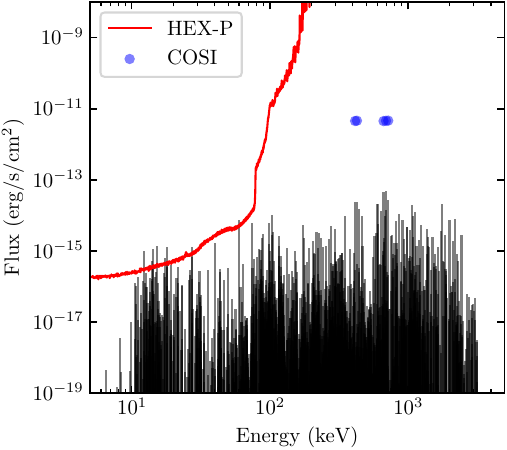}
    \caption{Synthetic spectra for $\unit[0.05]{M_{\odot}}$ of ejecta from a KNR $\unit[10]{kpc}$ away $\unit[10]{kyr}$ after the merger, plotted alongside 3$\sigma$ narrow-line sensitivities for COSI and HEX-P. The flux from this event is well below COSI's sensitivity, but HEX-P manages a marginal detection for an exposure time of $\unit[1]{Ms}$, assuming the remnant is $\lesssim\unit[1]{pc}$ in size (corresponding to an angular size comparable to HEX-P's HPD). A larger remnant would require a longer exposure time to maintain a detection significance of 3$\sigma$.}
    \label{fig:sensitivities}
\end{figure}

The High Energy X-ray Probe (HEX-P) is a proposed next-generation mission designed to achieve exceptional sensitivity across the $\unit[0.2-80]{keV}$ energy range \citep{2024FrASS..1157834M}. With a spectral resolution of $\sim 2\%$ FWHM at $\unit[20]{keV}$, HEX-P would be capable of resolving hard X-ray lines from r-process decay, especially for older remnants where doppler broadening is minimal. HEX-P's sensitivity would exceed that of COSI by orders of magnitude (see Fig. \ref{fig:sensitivities}), albeit over a dramatically smaller field of view: $11.3'\times 11.3'$ for the Low-Energy Telescope (LET), and $13.7' \times 13.7'$ for the High-Energy Telescope (HET). As such, a HEX-P-like instrument would not be suitable for all-sky surveys, but could fare better in focused observations of specific promising targets.

\begin{figure}
    \includegraphics{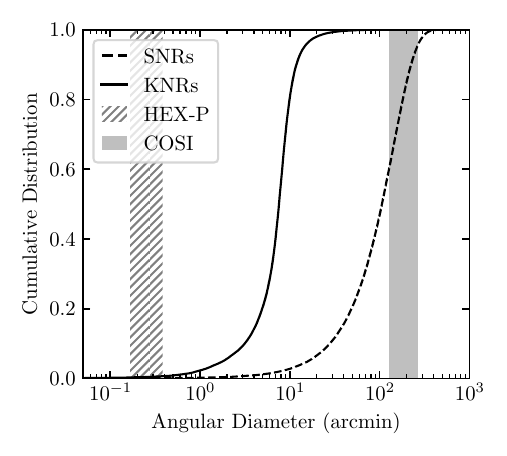}
    \caption{Angular diameters of KNRs and SNRs, shown alongside the angular resolutions (HPDs) of HEX-P and COSI. The shaded bands indicate the energy-dependent range of each instrument's angular resolution. For simplicity, only the dynamic orbital model with M.480 population synthesis data is shown for the KNRs. Nearly all KNRs and most SNRs appear point-like to COSI, whereas both KNRs and SNRs are highly extended relative to HEX-P's much finer angular resolution.}
    \label{fig:angular-sizes}
\end{figure}

The narrow-line sensitivity curve for HEX-P's HET assumes a point source smaller than the on-axis half-power diameter (HPD), $\sim 10"-20"$ across the energy range of interest. As shown in Fig. \ref{fig:angular-sizes}, the vast majority of KNRs and SNRs subtend angles that exceed this by two to three orders of magnitude. In the background-dominated limit, the signal-to-noise ratio scales as $\mathrm{SNR} \propto \sqrt{t\Omega_{\mathrm{src}}}$, where $t$ is the exposure time and $\Omega_{\mathrm{src}}$ is the angular area of the source in the sky. Thus, keeping a fixed $3\sigma$ significance while moving from a point source to an extended remnant inflates the required exposure time by a factor $\sim \Omega_{\mathrm{src}}/\Omega_{\mathrm{PSF}}$. This is problematic given that exposure times as long as $\sim\unit[1]{Ms}$ are already needed for marginal detections even under the most optimistic circumstances (e.g. Fig. \ref{fig:sensitivities}).

These stringent restrictions on exposure times make even small surveys (e.g. $\sim 1^{\circ}\times 1^{\circ}$) unfeasible. Mosaicking, each pointing would need to have a reduced exposure time to adhere to reasonable time allotment constraints. Consequently, a detection is effectively only possible under the rare circumstance in which a precise observing target has already been identified. While candidate KNRs have been suggested in the literature~\citep[e.g.][]{2019MNRAS.490L..21L}, another promising strategy may involve re-examining known SNRs that could possibly be misclassified KNRs. \citet{2019ApJ...880...23W} outline such a strategy and identify a number of nearby SNRs as potential targets, though all of the remnants listed are within a few kpc away and as such are extremely extended.

As it stands, in its current configuration, HEX-P remains an incisive line spectrograph but a poor survey instrument for old KNRs and SNRs. A natural next step would be a hybrid mission that combines HEX-P–class, high-spectral-resolution focusing optics with a co-aligned wide-field hard-X-ray module able to integrate the full flux of degree-scale remnants in a single exposure. Such a design removes the dominant solid-angle penalty that now forces multi-megasecond mosaics and would bring population studies of nearby KNRs and SNRs within realistic observing times.

The ESA M-class concept Advanced Surveyor for Transient Events and Nuclear Astrophysics (ASTENA) moves part-way toward this hybrid vision. Its $\unit[3]{m}$ Laue-lens Narrow Field Telescope (NFT) reaches deeper narrow-line sensitivity than HEX-P at $\gtrsim 70$ keV, yet the lens offers only a $\sim4'$ diameter FoV, which is around three times smaller than HEX-P’s, and a comparable $\sim30''$ PSF, so the same solid-angle handicap still applies for most remnants \citep{2021ExA....51.1203G,2021ExA....51.1175F}. ASTENA does attach a $\unit[2]{sr}$ coded-mask Wide-Field Monitor (WFM) that can encompass an entire remnant without mosaicking, but its $\sim4$–$6\%$ energy resolution and higher open-sky background leave its narrow-line sensitivity closer to COSI’s than to the order-of-magnitude gains that HEX-P’s focusing optics achieve.

Because near-term X-ray missions can only detect a KNR that is already pinpointed and compact, they cannot sample the full, degree-scale population predicted by our spatial-distribution model. The very requirement of an a-priori target means those same distribution predictions cannot, in turn, guide effective surveys with these instruments. Until a platform combines HEX-P–level spectroscopy with true wide-field coverage, the galactic distribution of KNRs will remain largely beyond observational reach via X-rays.

\section{Summary}
\label{sec:summary}

In this work, we set out to quantify how the Milky Way locations of r-process–enriched remnants from compact-binary mergers differ from those of core-collapse supernovae, and how these spatial differences translate into uncertainty in $\gamma$-ray detection prospects. SNRs were modeled with a five-arm logarithmic spiral distribution confined to the thin disk; their vertical scale height is $\sim\unit[100]{pc}$ and the radial profile falls exponentially with Galactocentric radius, consistent with observed high-energy SNR catalogues. By contrast, KNRs were assigned four alternative distributions that bracket the effects of merger kicks and long delay times: a disk model, static and dynamic orbital integrations in a Galactic potential, and an extreme isotropic case calibrated to short-GRB offsets. These prescriptions span mean radial offsets that differ by factors of a few and vertical dispersions that extend from the mid-plane to halo-scale heights.

When these spatial models are propagated through Monte-Carlo populations and compared with current instrument sensitivities, detectability proves vanishingly small. For a two-year all-sky survey with COSI, the expected number of detections ranges from a few $\times 10^{-4}-10^{-5}$ for KNRs, with a corresponding number of $\sim 10^{-4}$ for SNRs. Thus, the choice of spatial distribution introduces roughly an order-of-magnitude spread yet still leaves the absolute probability well below unity. Scaling the analysis to a COSI-like mission whose narrow-line sensitivity is improved by a factor $f$ shows that the detection fraction scales as $f^{1.388}$, and that model-to-model differences are most pronounced in the low-sensitivity regime and begin to vanish once $f \gtrsim 100$.

Hard-X-ray focusing telescopes such as the proposed HEX-P reach orders of magnitude deeper line sensitivity at $\sim \unit[20-80]{keV}$, but their sub-arcminute PSFs are mismatched to degree-scale remnants. Once the solid-angle penalty for such extended sources is included, a marginal $3\sigma$ detection requires multi-megasecond mosaics; in practice only a pre-identified, unusually compact target could be observed in a feasible exposure. Hybrid concepts that couple HEX-P–level spectroscopy to a wide-field hard-X-ray module would remove that limitation, but no such facility yet exists.

Thus, while spatial-distribution assumptions currently dominate the model-to-model spread in predicted fluxes, instrumental capability remains the fundamental bottleneck. A next-generation mission that preserves COSI’s all-sky reach but improves narrow-line sensitivity by roughly two orders of magnitude would shift the expected number of detectable remnants from effectively zero to order-unity, turning spatial-distribution differences into testable observational signatures.

\begin{acknowledgements}
    We would like to thank John Tomsick, Eric Burns, and Andreas Zoglauer for discussions on the detection capabilities of the COSI instrument. We would also like to thank Kaya Mori for sharing his knowledge of HEX-P and other hard X-ray missions, current and hypothetical. This work was supported by the US Department of Energy through the Los Alamos National Laboratory. Los Alamos National Laboratory is operated by Triad National Security, LLC, for the National Nuclear Security Administration of U.S.\ Department of Energy (Contract No.\ 89233218CNA000001).
\end{acknowledgements}

\appendix

\section{Time-Dependent Galactic Potential}
\label{sec:time_dependent_galactic_potential}

This appendix outlines the time-dependent galactic model used in our dynamic orbital integrations for KNRs. The model incorporates the evolving mass and scale lengths of the Milky Way's stellar disk, gas disk, and dark matter halo, allowing compact binary systems to orbit through a realistic time-varying gravitational potential.

\subsection{Stellar Disk Evolution}

We begin by modeling the growth of stellar mass $M_*$ through a delayed-$\tau$ star formation rate, following the form motivated by \citet{2014ApJS..214...15S} and \citet{2014ARA&A..52..415M}:
\begin{equation}
    \psi(t) = Ate^{-t/\tau}\,.
    \label{eqn:star-formation}
\end{equation}
Integrating this from $t'=0$ to $t'=t$, we obtain:
\begin{equation}
    M_*(t) = A\left[ \tau^2 - e^{-t/\tau}\left( \tau^2 + t\tau \right) \right]\,.
\end{equation}
We fix the present-day stellar mass at $\unit[6.08\times 10^{11}]{M_{\odot}}$ (as in Sec. \ref{subsec:disk-model}) and adopt $\psi_0=\unit[1.65]{M_{\odot} yr^{-1}}$ \citep{2015ApJ...806...96L}, yielding $A = \unit[5.238\times 10^9]{M_{\odot} \cdot Gyr^{-2}}$ and $\tau = \unit[3.612]{Gyr}$.

The time evolution of the stellar disk scale length $a_*$ follows the fit of \citet{2016ApJ...828...27N}, rescaled to produce the present-day local surface density \citet{2015ApJ...814...13M}. Assuming a solar orbital radius of $\unit[8.1]{kpc}$,
\begin{equation}
    a_*(t) = 1.405 \left[ \frac{M_*(t)}{\unit[10^{10}]{M_{\odot}}} \right]^{0.176}\unit[]{kpc}\,.
    \label{eqn:stellar-scalelength}
\end{equation}
The vertical scale height $b_*(t)$ is assumed proportional to $a_*(t)$, scaled to match the present-day local stellar volume density \citep{2015ApJ...814...13M}:
\begin{equation}
    b_*(t) = 0.197a_*(t)
\end{equation}

\subsection{Gas Disk Evolution}

We estimate the gaseous disk mass $M_{\mathrm{g}}(t)$ by inverting the star formation rate using the Kennicutt-Schmidt law \citep{1959ApJ...129..243S, 1989ApJ...344..685K, 1998ApJ...498..541K}, following the approach of \citet{2013ApJ...772..119L}:
\begin{equation}
    M_{\mathrm{g}}(t) = \left[ \frac{\psi(t)}{\alpha} \right]^{1/1.4}\,,
    \label{eqn:gaseous-mass}
\end{equation}
with star formation rate efficiency $\alpha = 9.668\times 10^{-6}$, calibrated to match the present-day gas mass from \citet{2006MNRAS.372.1149F}.

We assume gaseous disk scale length evolves proportionally to the stellar disk:
\begin{equation}
    a_{\rm g}(t) = 2.939a_*(t)\,,
    \label{eqn:gaseous-scalelength}
\end{equation}
rescaled to reproduce the present-day local gas surface density \citet{2015ApJ...814...13M}. The vertical scale height is:
\begin{equation}
    b_*(t) = 0.197 a_*(t)\,,
\end{equation}
chosen to match the present-day local gas volume density \citep{2015ApJ...814...13M}.

\subsection{Dark Matter Halo Evolution}

The dark matter halo mass is modeled as an exponential function of redshift:
\begin{equation}
    M_{\mathrm{DM}}(z) = M_{\mathrm{DM,0}}e^{-pz} \,,
\end{equation}
with $M_{\mathrm{DM},0} = \unit[7.855\times 10^{11}]{M_{\odot}}$ (see Sec. \ref{subsec:stat-orb-model}) and $p = 0.367$. The halo scale radius $r_s$ relates to the virial radius $r_{\mathrm{vir}}$ and concentration $c$ via:
\begin{equation}
    r_s = \frac{r_{\mathrm{vir}}}{c}, \hspace{0.25cm} \textrm{where} \hspace{0.25cm}  r_{\mathrm{vir}} = \left( \frac{3M_{\mathrm{DM}}}{4\pi\Delta \rho_c} \right)^{1/3}, \hspace{0.25cm} \Delta=200 \,.
\end{equation}
The critical density $\rho_c$ is defined as:
\begin{equation}
    \rho_c = \frac{3H^2}{8\pi G}, \hspace{0.25cm} H = H_0 \sqrt{\Omega_m (1 + z)^3 + \Omega_{\Lambda}} \,,
\end{equation}
with cosmological parameters $\Omega_m = 0.315$, $\Omega_{\Lambda} = 1 - \Omega_m = 0.685$, and $H_0 = \unit[67.4]{km \cdot s^{-1} \cdot Mpc^{-1}}$ \citep{2020A&A...641A...6P}.

The concentration evolves as:
\begin{equation}
    c = \frac{c_0}{1+z}, \hspace{0.25cm} c_0 = 24.384 \,,
    \label{eqn:concentration}
\end{equation}
which corresponds to $r_{s,0} = \unit[7.962]{kpc}$.

\section{Binary Population Properties}
\label{sec:binary-population-properties}

This appendix details the population synthesis ingredients required to assign birth times and progenitor metallicities to compact binary systems in the dynamic orbital model.

\subsection{Birth Time Distribution for Recent Mergers}

To accurately model KNRs formed within the last $\unit[10^6]{yr}$, we must determine the distribution of binary birth times that lead to mergers during this short observational window. This distribution, $\chi(t, t')$, gives the probability that a system born at time $t'$ merges between $t$ and $t + \Delta t$, where $t=(\unit[13.599]{Gyr})$ and $\Delta t = \unit[1]{Myr}$.

We compute this distribution by convolving the compact binary delay time distribution (DTD) with the star formation rate $\psi(t')$:
\begin{equation}
    \chi(t, t') \propto \displaystyle\int_t^{t+\Delta t} f(t')\textrm{DTD}(\tilde{t}-t')\psi(t') \hspace{0.1cm} d\tilde{t}\,,
\end{equation}
where $f(t')$ is the fraction of stars born at time $t'$ that eventually form merging compact binaries. Assuming $f(t')$ is constant and that the DTD scales as $\propto t^{-1}$ for $t > \unit[20]{Myr}$, this integral can be evaluated analytically for our delayed-$\tau$ star formation rate:
\begin{equation}
    \chi(t, t') \propto t'e^{-t' / \tau}\ln{\left| \frac{t + \Delta t - t'}{t - t'} \right|}\,,
\end{equation}
valid for $(t - t') \geq \unit[20]{Myr}$. This distribution features a broad component peaking near the epoch of maximal star formation, along with a sharp spike at recent times $t$ due to fast mergers. We sample from $\chi(t, t')$ to assign realistic birth times to compact binaries used in our orbital integrations.

\subsection{Redshift-Dependent Metallicity Assignment}

\begin{figure}
    \includegraphics[width=\linewidth]{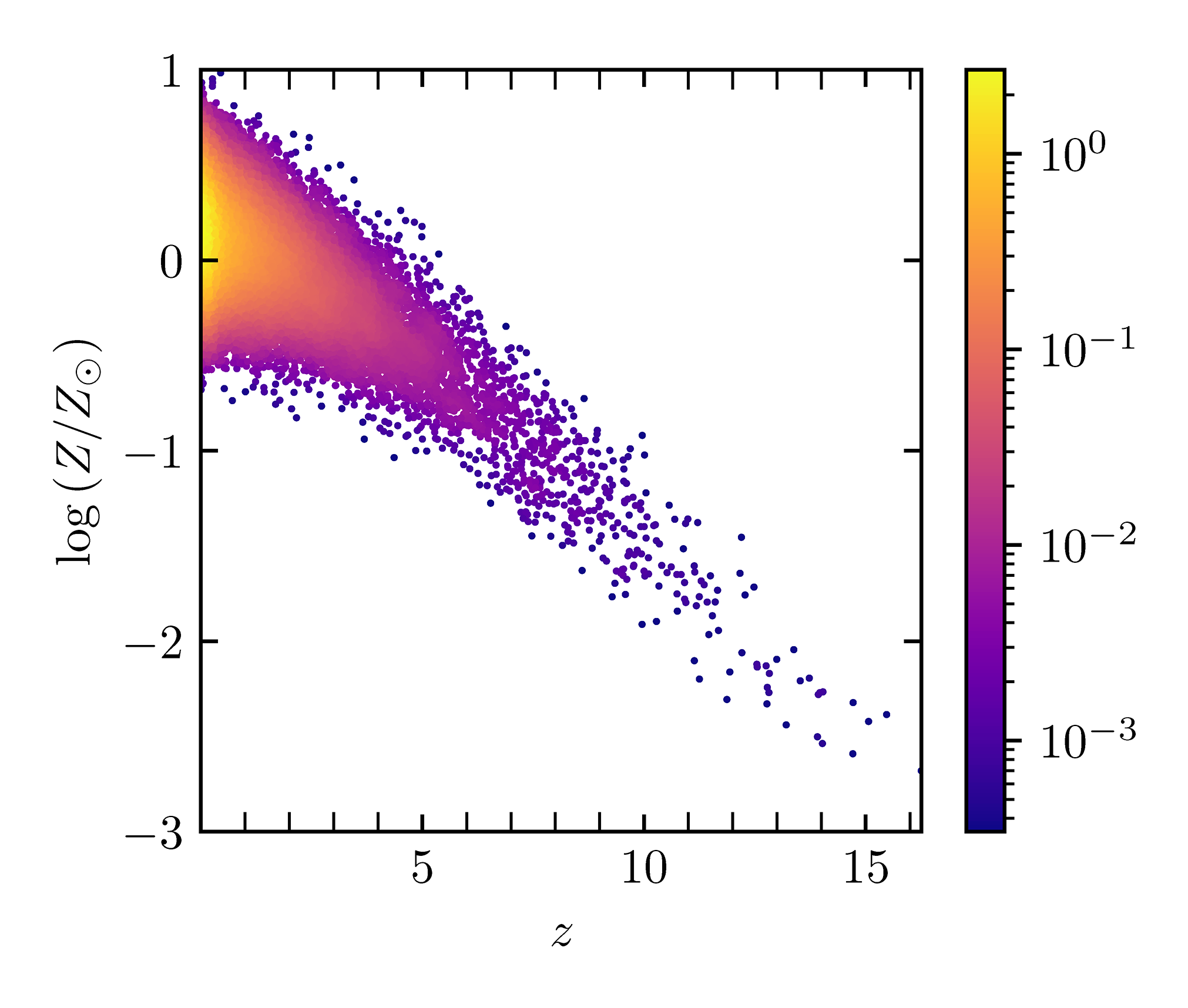}
    \caption{Redshift-metallicity relation used in our dynamic orbital model. Samples are drawn from a Gaussian distribution with a mean given by Eq. 6 in \citet{2017ApJ...840...39M} and a standard deviation of 0.2 dex.}
    \label{fig:redshift_metallicity}
\end{figure}

Merger properties such as kick velocities and delay times depend on progenitor metallicity, which in turn varies with redshift. To incorporate this, we assign a metallicity to each system based on its birth redshift using the empirical relation from \citet{2017ApJ...840...39M}:
\begin{equation}
    \log{\langle Z / Z_{\odot} \rangle} = 0.153 - 0.074z^{1.34}\,,
\end{equation}
with samples drawn from a Gaussian distribution centered on this mean and a standard deviation of $\sigma = \unit[0.2]{dex}$, consistent with observed scatter in the age-metallicity relation \citep{2004A&A...418..989N, 2007A&A...475..519H, 2001A&A...377..911F}. The resulting redshift-metallicity relation used in our model is shown in Fig. \ref{fig:redshift_metallicity}.

\section{Significant $\gamma$-producing isotopes}
\label{sec:gamma-producing-isotopes}

\begin{adjustwidth}{-.15in}{-.25in}
\begin{longtable*}{ll|cc|cc|l}
\caption{Potentially significant $\gamma$-ray emitters and their long-lived populating ancestors on timescales between 1 ky and 1 Gyr.  For each isotope, listed are its half-life $T_{1/2}$, several strongest $\gamma$-lines with their energies $E$ and intensities in photons ${\rm s}^{-1}\ {\rm g}^{-1}$, maximal isotopic mass fraction $X_{\rm max}$ in a typical $r$-process calculation, and a list of populating ancestor isotopes.  Photon fluxes for a line $\ell$ are estimated for a remnant at a distance $D = 3~{\rm kpc}$, assuming a total ejecta mass $m_{\rm ej} = 0.01~M_\odot$, according to the formula $\mathcal{F}_\ell = m_{\rm ej} X_{\rm max} I_\ell/(4\pi D^2)$.
} \\ 
\hline
Isotope & Halflife
& Line energy, $E$ & Intensity, $I$
& Mass fraction
& Line flux, $\mathcal{F}$
& Long-lived Ancestor(s)
\\
        & $T_{1/2}$
& [keV] & [ph s$^{-1}$ g$^{-1}$]
& $X_{\rm max}$
& [ph s$^{-1}$ cm$^{-2}$]
&
\\
\hline\hline
\endfirsthead
Isotope & $T_{1/2}$
& E [keV] & I [ph s$^{-1}$ g$^{-1}$]
& X$_{\rm max}$
& $\mathcal{F}$[ph s$^{-1}$ cm$^{-2}$]
& Long-lived Ancestor(s)
\\
\hline\hline
\endhead
\Ac{225}  &   9.920 d   &  10.6420 &  1.82$\times10^{14}$ & 4.80$\times10^{-9}$ & 1.55$\times10^{-8}$  & \Th{229},\U{233},\Np{237},\Cm{245}       \\
          &             &  12.0208 &  1.89$\times10^{14}$ &          & 1.61$\times10^{-8}$  &                                           \\
          &             &  14.7188 &  1.93$\times10^{14}$ &          & 1.65$\times10^{-8}$  &                                           \\
\Al{26}   &   717.0 ky  &  1808.65 &  7.05$\times10^{ 8}$ & 4.80$\times10^{-12}$ & 6.03$\times10^{-17}$  &  self \\
\Am{241}  &   432.6 y   &  13.9388 &  1.35$\times10^{10}$ & 2.03$\times10^{-4}$ & 4.88$\times10^{-8}$  &  \Cm{245} \\
          &             &  17.5393 &  1.33$\times10^{10}$ &          & 4.81$\times10^{-8}$  &                                           \\
          &             &  59.5409 &  4.54$\times10^{10}$ &          & 1.64$\times10^{-7}$  &                                           \\
\Am{243}  &   7.35  ky  &  74.6600 &  4.96$\times10^{ 9}$ & 1.26$\times10^{-3}$ & 1.11$\times10^{-7}$  &  \Cm{237},\Bk{247} \\
\Am{246}  &   39.0  m   &  14.9604 &  5.12$\times10^{17}$ & 2.42$\times10^{-12}$ & 2.20$\times10^{-8}$  &  \Cm{250} \\
          &             &  19.2323 &  6.28$\times10^{17}$ &          & 2.71$\times10^{-8}$  &  \\
          &             & 679.0000 &  4.25$\times10^{17}$ &          & 1.83$\times10^{-8}$  &  \\
\Bi{208}  &   358   ky  &2614.5    &  1.76$\times10^{ 8}$ & 1.00$\times10^{-15}$ & 3.13$\times10^{-21}$  &  self \\
\Bi{210}  &    3.04 My  & 265.6    &  1.05$\times10^{ 7}$ & 5.43$\times10^{-6}$ & 1.01$\times10^{-12}$  &  self \\
          &             & 304.6    &  5.78$\times10^{ 6}$ &          & 5.59$\times10^{-13}$  &  \\
          &             & 649.6    &  7.01$\times10^{ 5}$ &          & 6.78$\times10^{-14}$  &  \\
\Bi{213}  &   45.59 m   & 440.4500 &  1.85$\times10^{17}$ & 1.42$\times10^{-11}$ & 4.68$\times10^{-8}$  &  \Th{229},\U{233},\Np{237},\Cm{245} \\
\Bi{214}  &   19.71 m   & 609.3200 &  7.43$\times10^{17}$ & 3.43$\times10^{-12}$ & 4.54$\times10^{-8}$  & \Ra{226},\Th{230},\U{234},\Pu{242},\Cm{246},\Cm{250} \\
\Cm{245}  &   8.25  ky  &  14.2745 &  1.60$\times10^{ 9}$ & 3.73$\times10^{-3}$ & 1.06$\times10^{-7}$  & self \\
          &             &  18.0336 &  1.45$\times10^{ 9}$ &          & 9.64$\times10^{-8}$  & \\
          &             &  99.5232 &  1.19$\times10^{ 9}$ &          & 7.91$\times10^{-8}$  & \\
          &             & 103.7341 &  1.95$\times10^{ 9}$ &          & 1.29$\times10^{-7}$  & \\
\Cm{246}  &   4.706 ky  &  14.2737 &  3.38$\times10^{ 8}$ & 1.13$\times10^{-3}$ & 6.81$\times10^{-9}$  & self,\Cm{250} \\
          &             &  18.1426 &  3.74$\times10^{ 8}$ &          & 7.53$\times10^{-9}$  & \\
\Cm{247}  &   15.6  My  & 402.4000 &  2.47$\times10^{ 6}$ & 6.80$\times10^{-3}$ & 2.99$\times10^{-10}$  & self \\
\Cm{248}  &   348   ky  &  18.1425 &  4.01$\times10^{ 6}$ & 8.10$\times10^{-3}$ & 5.79$\times10^{-10}$  & self \\
\Fe{60}   &   2.62  My  &  58.603  &  1.73$\times10^{ 6}$ & 2.70$\times10^{-6}$ & 8.32$\times10^{-14}$  & self \\
\Fr{221}  &   4.9   m   & 218.1200 &  7.27$\times10^{17}$ & 1.58$\times10^{-12}$ & 2.04$\times10^{-8}$  &  \Th{229},\U{233},\Np{237},\Cm{245} \\
\Hf{182}  &   8.90  My  & 270.4080 &  6.43$\times10^{ 6}$ & 3.70$\times10^{-3}$ & 4.24$\times10^{-10}$  & self \\
\I{129}   &   16.14 My  &  29.7820 &  2.40$\times10^{ 6}$ & 2.50$\times10^{-2}$ & 1.07$\times10^{-9}$  & self, fission \\
\Kr{81}   &   229   ky  & 275.9900 &  2.11$\times10^{ 6}$ & 1.00$\times10^{-15}$ & 3.76$\times10^{-23}$  & self \\
\Mo{93}   &   4.0   ky  &  30.77   &  1.84$\times10^{ 5}$ & 1.00$\times10^{-15}$ & 3.28$\times10^{-24}$  & self \\
\Nb{92}   &   2.4   My  & 561.1    &  5.96$\times10^{ 7}$ & 1.00$\times10^{-15}$ & 1.06$\times10^{-21}$  & self \\
          &             & 934.5    &  4.41$\times10^{ 7}$ &          & 7.86$\times10^{-22}$  &  \\
\Nb{94}   &  20.4   ky  & 871.091  &  6.85$\times10^{ 9}$ & 1.56$\times10^{-11}$ & 1.90$\times10^{-15}$  & self \\
          &             & 702.65   &  6.86$\times10^{ 9}$ &          & 1.90$\times10^{-15}$  & \\
\Np{237}  &   2.144 My  &  16.6004 &  6.40$\times10^{ 6}$ & 1.42$\times10^{-2}$ & 1.62$\times10^{-9}$  & self,\Cm{245} \\
\Np{239}  &   2.356 d   & 103.7341 &  1.84$\times10^{15}$ & 1.09$\times10^{-9}$ & 3.57$\times10^{-8}$  & self,\Cm{247},\Bk{247} \\
          &             & 106.1230 &  2.24$\times10^{15}$ &          & 4.35$\times10^{-8}$  & \\
\Pa{231}  &   32.65 ky  &  12.7000 &  5.65$\times10^{ 8}$ & 2.46$\times10^{-3}$ & 2.47$\times10^{-8}$  &  self,\U{235},\Pu{239},\Am{243},\Cm{247},\Bk{247} \\
          &             &  27.36   &  1.83$\times10^{ 8}$ &          & 8.02$\times10^{-9}$  & \\
          &             & 300.066  &  4.21$\times10^{ 7}$ &          & 1.84$\times10^{-9}$  & \\
          &             & 302.667  &  4.01$\times10^{ 7}$ &          & 1.75$\times10^{-9}$  & \\
\Pa{233}  &   26.98 d   &  13.6000 &  4.10$\times10^{14}$ & 4.82$\times10^{-10}$ & 3.52$\times10^{-9}$  & \Np{237},\Cm{245} \\
\Pb{210}  &   22.20 y   &  10.8277 &  3.57$\times10^{11}$ & 1.99$\times10^{-6}$ & 1.26$\times10^{-8}$  & \Th{230},\U{234},\Pu{242},\Cm{246},\Cm{250} \\
          &             &  13.0050 &  2.77$\times10^{11}$ &          & 9.82$\times10^{-9}$  & \\
\Pb{211}  &   36.1  m   & 404.8530 &  3.43$\times10^{16}$ & 4.75$\times10^{-12}$ & 2.90$\times10^{-9}$  & \Pa{231},\U{235},\Pu{239},\Am{243},\Cm{247},\Bk{247}  \\
          &             & 832.0100 &  3.20$\times10^{16}$ &          & 2.71$\times10^{-9}$  & \\
\Pb{214}  &   27.06 m   & 351.9321 &  4.30$\times10^{17}$ & 4.67$\times10^{-12}$ & 3.58$\times10^{-8}$  &  \Ra{226},\Th{230},\U{234},\Pu{242},\Cm{246},\Cm{250} \\
\Pu{239}  &   24.11 ky  &  13.6082 &  3.38$\times10^{ 7}$ & 6.27$\times10^{-3}$ & 3.77$\times10^{-9}$  &  self,\Am{243},\Cm{247},\Bk{247} \\
\Pu{240}  &   6.561 ky  &  13.6085 &  3.62$\times10^{ 8}$ & 2.59$\times10^{-3}$ & 1.67$\times10^{-8}$  &  self,\Cm{248} \\
          &             &  17.1130 &  3.93$\times10^{ 8}$ &          & 1.81$\times10^{-8}$  &  \\
\Pu{246}  &   4.706 ky  &  14.6137 &  8.15$\times10^{ 7}$ & 3.96$\times10^{-3}$ & 5.75$\times10^{-9}$  &  \Cm{250} \\
          &             &  43.8100 &  1.10$\times10^{ 8}$ &          & 7.76$\times10^{-9}$  &  \\
          &             & 107.0430 &  8.99$\times10^{ 7}$ &          & 6.34$\times10^{-9}$  &  \\
          &             & 223.7500 &  1.03$\times10^{ 8}$ &          & 7.27$\times10^{-9}$  &  \\
\Ra{223}  &   11.43 d   &  81.3680 &  2.75$\times10^{14}$ & 2.28$\times10^{-9}$ & 1.11$\times10^{-8}$  &  \Pa{231},\U{235},\Pu{239},\Am{243},\Cm{247},\Bk{247} \\
          &             &  84.1450 &  4.51$\times10^{14}$ &          & 1.83$\times10^{-8}$  &  \\
          &             & 269.4630 &  2.63$\times10^{14}$ &          & 1.06$\times10^{-8}$  &  \\
\Ra{226}  &   1.600 ky  & 186.211  &  1.30$\times10^{ 9}$ & 4.67$\times10^{-3}$ & 1.08$\times10^{-7}$  & \Th{230},\U{234},\Pu{242},\Cm{246},\Cm{250} \\
\RRe{186} &   200   ky  &  58.009  &  6.41$\times10^{ 7}$ & 1.21$\times10^{-8}$ & 1.38$\times10^{-14}$  & self \\
          &             &  40.350  &  1.81$\times10^{ 7}$ &          & 3.90$\times10^{-15}$  &  \\
\Ra{225}  &   14.9  d   &  40.0000 &  4.32$\times10^{14}$ & 7.17$\times10^{-9}$ & 5.52$\times10^{-8}$  &  \Th{229},\U{233},\Np{237},\Cm{245} \\
\Ra{226}  &   1.600 ky  & 186.2110 &  1.31$\times10^{ 9}$ & 1.53$\times10^{-4}$ & 3.57$\times10^{-9}$  &  \Th{230},\U{234},\Pu{242},\Cm{246},\Cm{250} \\
\Sb{125}  &   2.758 y   &  27.4653 &  9.54$\times10^{12}$ & 3.75$\times10^{-7}$ & 6.37$\times10^{-8}$  &  fission \\
          &             & 427.8740 &  1.13$\times10^{13}$ &          & 7.55$\times10^{-8}$  &          \\
          &             & 600.5970 &  6.74$\times10^{12}$ &          & 4.50$\times10^{-8}$  &          \\
\Sb{126}  &   12.35 d   & 666.5000 &  3.08$\times10^{15}$ & 1.70$\times10^{-9}$ & 9.33$\times10^{-8}$  &  fission, \Sn{126} \\
          &             & 695.0000 &  3.08$\times10^{15}$ &          & 9.33$\times10^{-8}$  &          \\
\Sn{125}  &   9.64  d   &1089.1500 &  1.83$\times10^{14}$ & 3.59$\times10^{-9}$ & 1.17$\times10^{-8}$  &  fission \\
          &             &1067.1000 &  3.87$\times10^{14}$ &          & 2.47$\times10^{-8}$  &          \\
\Sn{126}  &   230.0 ky  &  87.5670 &  1.68$\times10^{ 8}$ & 1.16$\times10^{-2}$ & 3.47$\times10^{-8}$  &  self, fission \\
\Ta{182}  &   114.7 d   &  67.7497 &  9.84$\times10^{13}$ & 1.31$\times10^{-10}$ & 2.29$\times10^{-10}$  & fission, \Hf{182} \\
 \Tc{98}  &   4.2   My  & 745.35   &  3.20$\times10^{ 7}$ & 1.00$\times10^{-15}$ & 5.70$\times10^{-22}$  & self, fission \\
          &             & 652.41   &  3.20$\times10^{ 7}$ &          & 8.04$\times10^{-10}$  &  \\
 \Tc{99}  &   211.0 ky  &  89.5    &  4.41$\times10^{ 3}$ & 1.41$\times10^{-3}$ & 1.10$\times10^{-13}$  & self, fission \\
\Th{227}  &   18.70 d   &  50.1300 &  1.29$\times10^{14}$ & 3.75$\times10^{-9}$ & 8.62$\times10^{-9}$  &  \Pa{231},\U{235},\Pu{239},\Am{243},\Cm{247},\Bk{247} \\
          &             & 235.9600 &  1.99$\times10^{14}$ &          & 1.33$\times10^{-8}$  &  \\
\Th{229}  &   7.916 ky  &  12.3318 &  4.26$\times10^{ 9}$ & 1.42$\times10^{-3}$ & 1.07$\times10^{-7}$  &  self,\U{233},\Np{237},\Cm{245} \\
          &             &  15.1900 &  4.83$\times10^{ 9}$ &          & 1.22$\times10^{-7}$  &  \\
          &             &  11.1    &  8.94$\times10^{ 8}$ &          & 2.26$\times10^{-8}$  &  \\
          &             & 193.52   &  3.20$\times10^{ 8}$ &          & 8.10$\times10^{-9}$  &  \\
          &             & 210.853  &  2.03$\times10^{ 8}$ &          & 5.14$\times10^{-9}$  &  \\
\Th{230}  &   75.4  ky  &  12.3000 &  5.81$\times10^{ 7}$ & 4.25$\times10^{-3}$ & 4.40$\times10^{-9}$  &  self,\U{234},\Pu{242},\Cm{246},\Cm{250} \\
          &             &  67.672  &  2.88$\times10^{ 6}$ &          & 2.18$\times10^{-10}$  &  \\
 \U{233}  &   159.2 ky  &  12.9597 &  8.36$\times10^{ 6}$ & 3.92$\times10^{-3}$ & 5.84$\times10^{-10}$  &  self,\Np{237},\Cm{245} \\
          &             &  16.0978 &  8.69$\times10^{ 6}$ &          & 6.07$\times10^{-10}$  &  \\
          &             &  42.4349 &  2.55$\times10^{ 5}$ &          & 1.78$\times10^{-11}$  &  \\
          &             &  97.1346 &  7.21$\times10^{ 4}$ &          & 5.04$\times10^{-12}$  &  \\
          &             &  54.7039 &  5.96$\times10^{ 4}$ &          & 4.16$\times10^{-12}$  &  \\
 \U{234}  &   245.5 ky  &  16.1500 &  1.17$\times10^{ 7}$ & 6.93$\times10^{-3}$ & 1.44$\times10^{-9}$  &  self,\Pu{242},\Cm{246},\Cm{250} \\
          &             &  53.20   &  2.82$\times10^{ 5}$ &          & 3.48$\times10^{-11}$  &  \\
          &             & 120.90   &  7.84$\times10^{ 4}$ &          & 9.68$\times10^{-12}$  &  \\
 \U{235}  &   1.0   My  & 185.7150 &  1.20$\times10^{ 5}$ & 5.85$\times10^{-3}$ & 1.25$\times10^{-11}$  &  self,\Pu{239},\Am{243},\Cm{247},\Bk{247} \\
 \U{236}  &   23.42 My  &  16.1501 &  1.29$\times10^{ 5}$ & 1.37$\times10^{-2}$ & 3.15$\times10^{-11}$  &  self,\Np{226},\Pu{240},\Cm{248}
 \\
\hline
\label{tab:gammas}
\end{longtable*}
\end{adjustwidth}

\bibliography{bibliography}{}
\bibliographystyle{aasjournal}

\end{document}